\documentclass[aos,preprint]{imsart}
\setattribute{journal}{name}{}
\RequirePackage[OT1]{fontenc}
\RequirePackage{amsthm,amsmath}
\RequirePackage[numbers]{natbib}
\usepackage{amsmath,amsfonts,amssymb,amsthm,mathrsfs}
\usepackage{algorithm} 
\usepackage{algorithmicx}
\usepackage{algpseudocode}
\usepackage{tikz}
\usetikzlibrary{chains,matrix,scopes,fit,decorations,positioning,shapes,arrows,decorations.markings}
\startlocaldefs
\numberwithin{equation}{section}
\newtheorem{thm}{Theorem}[section]
\newtheorem{lem}[thm]{Lemma}
\newtheorem{prop}[thm]{Proposition}
\newtheorem{cor}[thm]{Corollary}
\theoremstyle{definition}
\newtheorem{defn}[thm]{Definition}

\newtheorem{ex}[thm]{Example}
\newtheorem{assp}{Assumption}
\newtheorem{rem}[thm]{Remark}

\DeclareMathOperator{\Supp}{supp}

\DeclareMathOperator{\tr}{tr}

\newcommand{\cA}{\mathcal{A}}

\newcommand{\cC}{\mathcal{C}}

\newcommand{\E}{\mathbb{E}}
\newcommand{\cE}{\mathcal{E}}

\newcommand{\cK}{\mathcal{K}}

\renewcommand{\P}{\mathbb{P}}

\newcommand{\R}{\mathbb{R}}
\renewcommand{\S}{\mathbb{S}}

\newcommand{\cX}{\mathcal{X}}

\newcommand{\mtp}{{\rm MTP}_{2}}

\newcommand{\cd}{\,|\,}
\newcommand{\cip}{\mbox{\,$\perp\!\!\!\perp$\,}}
\newcommand{\indep}{\cip}

\newcommand{\pto}{{\mathcal{P}_2}}
\newcommand{\spa}{\mathcal{X}}

\endlocaldefs

\begin{document}

\begin{frontmatter}
\title{Total positivity in exponential families with application to binary variables}
\runtitle{Total positivity for binary variables}

\begin{aug}
\author{\fnms{Steffen} \snm{Lauritzen}\thanksref{m1}\ead[label=e1]{lauritzen@math.ku.dk}},
\author{\fnms{Caroline} \snm{Uhler}\thanksref{m2}\ead[label=e2]{cuhler@mit.edu}}
\and
\author{\fnms{Piotr} \snm{Zwiernik}\thanksref{m3}
\ead[label=e3]{piotr.zwiernik@upf.edu}}

\runauthor{S. Lauritzen et al.}

\affiliation{University of Copenhagen\thanksmark{m1}, Massachusetts Institute of Technology\thanksmark{m2}, \\ETH Zurich\thanksmark{m2} and Universitat Pompeu Fabra\thanksmark{m3}}

\address{Department of Mathematical Sciences\\ University of Copenhagen\\ Copenhagen, Denmark\\
\printead{e1}}
\address{Laboratory for Information and Decision Systems, \\and Institute for Data, 
Systems, and Society\\ Massachusetts Institute of Technology\\ Cambridge, MA, USA\\
\printead{e2}}

\address{Department of Economics and Business\\ Universitat Pompeu Fabra\\ Barcelona, Spain\\
\printead{e3}}
\end{aug}

\begin{abstract}
We study exponential families of distributions that are multivariate totally positive of order $2$ ($\mtp$), show that these are convex exponential families, and derive conditions for existence of the MLE. Quadratic exponential familes of $\mtp$ distributions contain attractive Gaussian graphical models and ferromagnetic Ising models as special examples. We show that these are defined by intersecting the space of canonical parameters with a polyhedral cone whose faces correspond to conditional independence relations. Hence $\mtp$ serves as an implicit regularizer for quadratic exponential families and leads to sparsity in the estimated graphical model. We prove that the maximum likelihood estimator (MLE) in an $\mtp$ binary exponential family exists if and only if both of the sign patterns $(1,-1)$ and $(-1,1)$ are represented in the sample for every pair of variables; in particular, this implies that the MLE may exist with $n=d$ observations, in stark contrast to unrestricted binary exponential families where $2^d$ observations are required.  Finally, we provide a novel and globally convergent algorithm for computing the MLE for $\mtp$ Ising models similar to iterative proportional scaling and apply it to the analysis of data from two psychological disorders. 
\end{abstract}



\end{frontmatter}

\section{Introduction and motivation}

This paper discusses exponential families and, in particular, binary graphical models with a special form of positive dependence. Total positivity is a  strong form of positive dependence that has become an important concept in modern statistics; see, e.g., \cite{colangelo2005some,KarlinRinott80}. This property (also called the $\mtp$ property) appeared in the study of stochastic orderings, asymptotic statistics, and in statistical physics \cite{fortuin1971correlation,newman1983general}. Families of distributions with this property lead to many computational advantages \cite{bartolucci2002recursive,djolonga2015scalable,propp1996exact} and they are a convenient shape constraint in nonparametric statistics \cite{robeva2018maximum}. They also became a useful tool in modelling with latent variables; see \cite{forcina2000} for an overview. In particular, in \cite{ARSZ}  the $\mtp$ property explicitly appeared in the description of the binary~latent~class~model.

In the Gaussian setting, the $\mtp$ property  was shown to simplify inference~\cite{anandkumar2012high,malioutov2006walk}. In this case the $\mtp$ property is equivalent to the covariance matrix being an inverse M-matrix, which is a linear constraint on the concentration matrix. This led Slawski and Hein~\cite{slawski2015estimation} to propose efficient learning procedures based on convex optimization; see also \cite{bhattacharya2012covariance,egilmez2017graph,LUZ}. The present paper develops similar results for  exponential families with special emphasis on models  for binary variables, including ferromagnetic Ising models.
Our main results are the following: \begin{itemize}
\item We show in Section~\ref{sec:expfamilies} that the $\mtp$ property is given by a convex constraint in an exponential family and use convex optimization theory to derive necessary and sufficient conditions ensuring than an estimate maximizes the likelihood. For a quadratic exponential family, including the Ising model for binary variables, the KKT conditions yield sparsity in the associated matrix for interaction potentials. 
\item We show in
Section~\ref{sec:binary} that the KKT conditions ensure context-dependent conditional independence restrictions
and that for binary variables  the MLE exists under $\mtp$ if and only if both of the sign patterns $(1,-1)$ and $(-1,1)$ are represented in the sample for every pair of variables. This ensures the minimal sample size for the MLE to exist be of order $d$ rather than $2^d$ where $d$ is the number of variables considered.
\item We show --- also  in Section~\ref{sec:binary} --- that adding conditional independence assumptions by further assuming a graphical model, reduces this condition to hold for pairs of vertices $ij$ that are neighbours in the graph, reducing the order of the minimal sample size to be the maximal clique size of the graph. 
\item We show --- also  in Section~\ref{sec:binary} --- that for symmetric binary $\mtp$ distributions, including ferromagnetic Ising models with no external field, presence of just one of the sign patterns $(1,-1)$ and $(-1,1)$ for every pair ensures existence of the MLE;
\item We develop --- in Section~\ref{sec:ising} --- a novel IPS type algorithm for calculating the MLE in a ferromagnetic Ising model that is shown to be globally convergent.
\end{itemize}

The remainder of this paper is structured as follows: In Section~\ref{sec:notation} we formally introduce $\mtp$ distributions and associated notation. 

In Section~\ref{sec:application} we apply our results to the analysis of two psychological disorders, showing that the resulting $\mtp$ graphical model is highly interpretable and consistent with domain knowledge.

\section{Preliminaries}\label{sec:notation}

Let $V=\{1,\dots , d\}$ be a finite set and let $X=(X_v,v\in V)$ be random variables with labels in $V$. 
We consider the product space $\cX=\prod_{v\in V}\cX_{v}$, where $\cX_v\subseteq \R$ is the state space of $X_v$, inheriting the order from $\R$.
In this paper, the state spaces are either discrete (finite sets) or open intervals on the real line. 

\begin{assp}
	All distributions are assumed to have densities with respect to the product measure $\mu=\otimes_{v\in V}\mu_v$, referred to as the \emph{base measure},  where $\mu_v$ is the counting measure if $\cX_{v}$ is discrete, and $\mu_v$ is the Lebesgue measure giving length 1 to the unit interval if $\cX_{v}$ is an open interval. 
\end{assp}
 We note that any other equivalent product measure can be used as base measure without affecting the $\mtp$ property as defined below.

A function $f$ on $\cX$ is said to be \emph{multivariate totally positive of order $2$} ($\mtp$) if 
\begin{equation}\label{eq:MTP2}
f(x)f(y)\quad\leq \quad f(x\wedge y)f(x\vee y)\qquad\mbox{for all }x,y\in \cX,
\end{equation}
where $x\wedge y$ and $x\vee y$ denote the elementwise minimum and maximum, i.e., 
$$x\wedge y=(\min(x_{v},y_{v}),v\in V),\quad x\vee y=(\max(x_{v},y_{v}),v\in V).$$

These inequalities are non-trivial only if $x,y\in \cX$ are \emph{not comparable}, that is, neither $x\leq y$ nor $x\geq y$. For $d=2$, a function that is $\mtp$ is simply called \emph{totally positive}~\cite{KarlinRinott80}.  We say that $X$ or the distribution of $X$ is $\mtp$ if its density function $p$ is $\mtp$. 

For strictly positive distributions, $\mtp$ can be verified by checking that (\ref{eq:MTP2}) holds for $x,y\in \cX$ that are not comparable and differ in exactly two coordinates; c.f. \cite[Proposition~2.1]{KarlinRinott80}. We call such pairs \emph{elementary} and denote the set of all elementary pairs by $\cE\subset \cX\times \cX$. For more details on $\mtp$ distributions, see \cite{KarlinRinott80} and \cite{fallat:etal:17}.

\section{Totally positive exponential families}\label{sec:expfamilies}

We first consider $\mtp$ for exponential families and show that maximum likelihood estimation for exponential families under $\mtp$ leads to a convex optimization problem. We then discuss conditions for the existence of the MLE and finally specialize these results to quadratic exponential families, which include as prominent examples the Gaussian distribution and the Ising model.

\subsection{Convexity of totally positive exponential families}
Consider an exponential family with density $p(x;\theta)$ satisfying
\begin{equation}\label{eq:expfam}
\log p(x;\theta)=\langle\theta,T(x)\rangle-A(\theta)+g(x),	
\end{equation}
with sample space $\mathcal{X}$, sufficient statistics $T:\cX\to \R^k$ and base measure $\mu$. Assume that the family is \emph{minimally represented}, i.e.\ that $\langle\lambda,T(X)\rangle +b=0$ almost surely implies $\lambda=0$, and that the family is \emph{regular} so that 
the space of canonical parameters
$$\mathcal{K}=\{\theta\in\mathbb{R}^k \; :\; A(\theta)<\infty\}$$
is an open convex set. 
\begin{assp}\label{as:as1}
	Throughout, we assume that there exists $\theta_0$ such that $p(x;\theta_0)$ is a product distribution, or equivalently,
\begin{equation}\label{eq:modular}
p(x\vee y;\theta_0)p(x\wedge y;\theta_0)\;=\;p(x;\theta_0)p(y;\theta_0)\qquad\mbox{for all }x,y\in \cX.	
\end{equation}
Since every distribution in an exponential family can act as the base distribution, we can then pick $p(x;\theta_0)$ as the base measure. It then holds that
$$
g(x\vee y)+g(x\wedge y)-g(x)-g(y)\;=\;0.
$$
We say that such an exponential family \textit{has a product base}. \end{assp}

  All exponential families that contain a full independence distribution admit a product base. This includes all models discussed in this article and in particular Gaussian graphical models and log-linear models.

For an exponential family of the form (\ref{eq:expfam}) and any two $x,y\in \cX$ we define
$$
\Delta(x,y;\theta)\;\;:=\;\;\log \left(\frac{p(x\vee y;\theta)p(x\wedge y;\theta)}{p(x;\theta)p( y;\theta)}\right).
$$
The density $p(x;\theta)$ is $\mtp$ if and only if $\Delta(x,y;\theta)\geq 0$ for all elementary pairs in $\cE$. For exponential families with a product base it holds that
$$
\Delta(x,y;\theta)\;=\;\big\langle\theta,T(x\wedge y)+T(x\vee y)-T(x)-T(y)\big\rangle,
$$
which is an \emph{affine} function in $\theta$. \begin{defn}\label{def:K2}
The set $\cK_2\subset \cK$ of totally positive canonical parameters is the subset of canonical parameters for which the density $p(x;\theta)$ is $\mtp$.
\end{defn}
Since $\cK_2$ is given by the linear inequalities $\Delta(x,y;\theta)\geq 0$ for all $x,y\in \cX$, we immediately get the following result.

\begin{thm}
	\label{th_conv_exp_fam}
The $\mathcal{K}_2$ of totally positive canonical parameters is a convex set that is relatively closed in~$\cK$. 
\end{thm}
We note that this result holds also for exponential families without a product base. However, in that case the set of $\mtp$ canonical parameters $\mathcal{K}_2$ may be empty. 

In \cite{LUZ} we considered the Gaussian setting and showed that $\mathcal{K}_2$ is a convex cone. By essentially the same argument, this extends to discrete Gaussian distributions over $\cX=\mathbb Z^d$, which were introduced in~\cite{agostini2019discrete}. More generally, we obtain the following result. 

\begin{prop}\label{prop:generateC}
	The set $\cK_2\,$ is obtained by intersecting $\cK$ with a closed convex cone $\cC\subseteq \R^d$, whose dual cone is the closure of the cone generated by the set
	$$\{T(x\wedge y)+T(x\vee y)-T(x)-T(y):\; x,y\in\mathcal{E}\}.$$
	
\end{prop}

\begin{proof} The set of inequalities $\Delta(x,y;\theta)\geq 0$, one for each elementary pair $x,y\in\mathcal{E}$, defines a convex cone in $\theta\in\mathbb{R^d}$. We have $\langle \theta,T(x\wedge y)+T(x\vee y)-T(x)-T(y)\rangle\geq 0$ for all $x,y\in\mathcal{E}$ if and only if $\langle \theta,v\rangle\geq 0$ for all $v$ in the cone generated by the set $\{T(x\wedge y)+T(x\vee y)-T(x)-T(y)\;:\; x,y\in\mathcal{E}\}$; denote this cone by $\cC^*$. This shows that $\cC=(\cC^*)^\vee$ and so $\cC^\vee=(\cC^*)^{\vee\vee}$. The latter is equal to the closure of $\cC^*$ by the standard theory of convex cones; see, for example, \cite[Section 2.6.1]{boyd}.  
\end{proof}

\begin{rem}
	\label{remark_polyhedral}
	When $\cX$ is finite, i.e.\ for log-linear models, Proposition~\ref{prop:generateC} implies that $\cC$ is polyhedral. Since $\cC$ is polyhedral also in the Gaussian setting, finiteness of $\cX$ is not a necessary condition. In fact, we will show in Proposition~\ref{prop:mtpquadratic} that $\cC$ is polyhedral for any quadratic exponential family. When $\cC$ is polyhedral, then every face of $\cC$ intersected with $\cK$ corresponds to the $\mtp$ distributions in an exponential subfamily.
\end{rem}

\subsection{The MLE and its existence}
An important consequence of Theorem~\ref{th_conv_exp_fam} is that any $\mtp$ exponential family is a convex exponential family and thus the maximum likelihood estimator (MLE), if it exists, is uniquely defined; see \cite[Section 9.4]{barndorff:78}.

Let $U=\{x^1,\ldots,x^n\}$ denote a  sample of size $n$ and let $\bar T:=\frac{1}{n}\sum_{i}T(x^i)$ be the average of the corresponding sufficient statistics. Let $\mathcal{S}$ denote the interior of $\textrm{conv}({\rm supp}(\mu\circ T^{-1}))$, the convex support of the sufficient statistics. Then by the general theory of exponential families~\cite{barndorff:78}, the MLE $\hat\theta$ exists if and only if $\bar T$ lies in $\mathcal{S}$, in which case it is \emph{uniquely} defined by 
$$
\nabla A(\hat \theta)\;\;=\;\;\E_{\hat\theta} [T(X)]\;\;=\;\; \bar T.
$$ 
The following theorem extends this result to a characterization of existence of the MLE for the subfamily of $\mtp$ distributions. By Proposition~\ref{prop:generateC} there exists a closed convex cone $\mathcal{C}$ such that the space of all $\mtp$ canonical parameters is given by $\mathcal{K}_2 = \mathcal{K}\cap \mathcal{C}$. We define
$$\mathcal{S}_2 := \mathcal{S}-\mathcal{C}^\vee$$
as the Minkowski sum of $\mathcal{S}$ with the dual of $-\mathcal{C}$; c.f. Proposition \ref{prop:generateC}.

\begin{thm}
\label{thm_mtp_exp_family}
Let $p(x;\theta)$ be a minimally represented regular exponential family. Then the MLE $\hat\theta$ based on $\bar T$ exists in the $\mtp$ submodel if and only if $\bar T\in\mathcal{S}_2$, in which case $\hat\theta$ is uniquely defined by
\begin{enumerate}
	\item[(a)] primal feasibility: \;$\hat\theta\in\mathcal{K}_2$,
	\item[(b)] dual feasibility: \quad$ \hat{\sigma}:=\nabla A (\hat{\theta})\in\mathcal{S}$ with $\hat\sigma-\bar T \in\mathcal{C}^\vee$,
	
	\item[(c)] complementary slackness: \;\; $\langle \hat\theta,\, \hat\sigma-\bar T\rangle= 0$.
\end{enumerate}
\end{thm}

\begin{proof} The maximum likelihood estimation problem can be formulated as the following optimization problem:
	\begin{equation*}
	\begin{aligned}
	& \underset{\theta\in\mathcal{K}}{\text{maximize}}
	& & \langle \theta, \bar T\rangle - A(\theta) \\
	& \text{subject to} && \theta\in\mathcal{C}.
	\end{aligned}
	\end{equation*}
	This is a convex optimization problem, since $A(\theta)$ is convex on $\mathcal{K}$. The Lagrangian is
	$$\mathcal{L}(\theta,\lambda) =  \langle \theta, \bar T\rangle - A(\theta) + \langle \theta,\lambda \rangle,$$
	where $\lambda\in \cC^\vee$. Let $A^*$ denote the conjugate dual of $A$ with domain $\mathcal{S}$. Then 
	$$\max_{\theta\in\mathcal{K}} \mathcal{L}(\theta,\lambda) = A^*(\bar T+\lambda),$$
	and hence the dual optimization problem is given by
	\begin{equation*}
	\begin{aligned}
	& \underset{\sigma\in\mathcal{S}}{\text{minimize}}
	& & A^*(\sigma)  \\
	& \text{subject to}
	&& \sigma-\bar T\in\mathcal{C}^\vee.
	\end{aligned}
	\end{equation*}
	The MLE exists if and only if the primal and dual problems are feasible. The primal problem is feasible by the assumption $\mathcal{K}_2\neq\emptyset$. The dual problem is feasible if and only if $\bar T\in\mathcal{S}_2$. The characterization of the MLE then follows from the KKT conditions.
\end{proof}

As in the Gaussian case, complimentary slackness imposes sparsity in the MLE $\hat\theta$. This property makes $\mtp$ exponential families potentially useful in high dimensional contexts. Before we discuss this in further detail, we shall consider the case of a quadratic exponential family, including the Gaussian case and Ising models.

\subsection{Quadratic exponential families}

The density function of a \emph{quadratic exponential family}  is of the form
\begin{equation}\label{eq:quadratic}
p(x; h,J)\;\;=\;\;\exp\left(h^Tx+x^T J  x/2-A(h,J )\right),	
\end{equation}
with $h\in \R^d$ and $J \in \S^d$, where $\S^d$ is the set of symmetric matrices in $\R^{d\times d}$ so here the canonical parameter space is $\mathcal{K}=\R^d\times \S^d$.  Important examples of such exponential families in the discrete setting are Ising models, which we discuss in more detail in Section \ref{sec:ising}, and Gaussian graphical models in the continuous setting. Note that in the binary setting we require $J _{ii}=0$ in order to obtain a minimally represented exponential family. We start by showing that $\mathcal{C}$ is a polyhedral cone for any quadratic exponential family. 
\begin{prop}\label{prop:mtpquadratic}
	The subfamily of $\mtp$ distributions in a quadratic exponential family is obtained by intersecting $\mathcal{K}$ with a polyhedral cone $\,\mathcal{C}$, namely the cone $\S^d_+=\{J \in \S^d \mid J _{ij}\geq 0 \textrm{ for all } i\neq j\}$.
\end{prop}
\begin{proof}
	By \cite[Theorem 7.5]{fallat:etal:17}, a quadratic exponential family is $\mtp$ if and only if $\exp(J _{ij}x_ix_j)$ is $\mtp$ for all $i\neq j$. This is the case if and only if for every $x,y$ that differ in two coordinates $i,j$ with $x_i<y_i$ and $x_j>y_j$, it holds that
	$$
	J _{ij}(y_i-x_i)(x_j-y_j)\geq 0,
	$$
	or equivalently $J _{ij}\geq 0$. This completes the proof. 
\end{proof}

We denote the mean parameters by $\mu:=\E_\theta X$ and $\Xi:=\E_\theta XX^T$. Then $(\mu,\Xi)$ can be transformed to $(\mu,\Sigma)$, where $\Sigma=\Xi-\mu \mu^T$ is the covariance matrix of $X$. Note that then
$$
\cC=\{(h,J )\in \R^d\times \mathbb S^d:\,J _{ij}\geq 0\mbox{ for } i\neq j\}.
$$
Each facet of $\cC$ corresponds to one of the $J_{ij}$'s being zero; c.f. Remark~\ref{remark_polyhedral}. Equivalently, by the Hammersley-Clifford theorem, each facet consists of members in the $\mtp$ exponential family that satisfy the conditional independence relation $X_i\indep X_j|X_{V\setminus \{i,j\}}$. The dual cone of $\cC$ is given by
\begin{equation}\label{eq:Cveequadr}
\cC^\vee=\{(0,\Xi )\in \R^d\times \mathbb S^d:\, \Xi _{ij}\geq 0\mbox{ for } i\neq j, \textrm{ and } \,\Xi_{ii}=0\mbox{ for all }i\}.	
\end{equation}

Let $U=\{x^1,\ldots,x^n\}$ as before be a sample of size $n$ and let $\bar x=\frac{1}{n}\sum_i x^i$ and $M=\frac{1}{n}\sum_i x^i (x^i)^T$ be the corresponding sample averages. Let $S=M-\bar x\bar x^T$ denote the sample covariance matrix. By standard exponential family theory, the MLE in the quadratic exponential family \eqref{eq:quadratic} corresponds to the unique distribution in the family which matches the sample averages, i.e., $(\hat{\mu},\hat\Xi)=(\bar{x},M)$, or equivalently, $(\hat{\mu},\hat\Sigma)=(\bar{x},S)$. By adding the $\mtp$ constraint, the situation changes somewhat. As a direct corollary to Theorem~\ref{thm_mtp_exp_family} we obtain the following result regarding the MLE in an $\mtp$ quadratic exponential family. 

\begin{cor}\label{cor_MLE_MTP2}
	Let $p(x;h,J)$ be a minimal regular quadratic exponential family. Let $\bar{x}$ and $S$ be the sample mean and covariance matrix. Then the corresponding MLE $(\hat h,\hat J )\in \cK$ with $(\hat \mu,\hat \Xi ):=\nabla A(\hat h,\hat J )$ and $\hat\Sigma:=\hat\Xi-\hat\mu\hat\mu^T$,  is uniquely defined by
	\begin{itemize}
		\item[(i)] $\hat J _{ij}\geq 0$ for $i\neq j$,
		\item[(ii)] $\hat \mu=\bar x$, $\hat \Sigma_{ii}=S_{ii}$,  and $\hat \Sigma_{ij}\geq S_{ij}$ for $i\neq j$,
		\item[(iii)]  $(\hat \Sigma_{ij}-S_{ij})\hat J _{ij}=0$ for all $i\neq j$. 
	\end{itemize}
\end{cor}
\begin{proof}
	The conditions of Theorem~\ref{thm_mtp_exp_family} translate precisely to (i), (ii), (iii), namely the primal feasibility condition is derived in Proposition~\ref{prop:mtpquadratic}, the dual feasibility condition follows from \eqref{eq:Cveequadr}, 
	and the complementary slackness condition follows from the fact that the inner product between dual cones is zero if and only if each summand is zero.
\end{proof}
\begin{rem}In quadratic exponential families the condition $\overline T\in \mathcal S_2=\mathcal S-\mathcal C^\vee$, that assures existence of the MLE, translates to the condition~(ii) in Corollary~\ref{cor_MLE_MTP2}. This condition can again be expressed more explicitly in terms of the observations: in the Gaussian case this becomes equivalent to all correlations being numerically less than one (\cite{LUZ}), and we derive the explicit conditions for our cases in  Theorem~\ref{thm:fullMLE}, Corollary~\ref{cor:fullMLE}, and Theorem~\ref{th:mainMLE}.
\end{rem}
\begin{rem} Note also  that in the binary case, where we have $J_{ii}=0$ and $\Xi_{ii}=1$ for all $i$, the condition (ii) reduces to 
$\hat \mu=\bar x$,  and $\hat \Sigma_{ij}\geq S_{ij}$ for $i\neq j$. 
\end{rem}
The specialization of this result to Gaussian graphical models was discussed in~\cite{LUZ}. Note that the $\mtp$ constraint induces sparsity in the MLE  $\hat J$ through the complementary slackness constraint (iii). For example, if $S_{ij}<0$, then complementary slackness implies that $\hat J_{ij}=0$ simply because in an $\mtp$ distribution all covariances are positive. The sparsity pattern of $\hat{J}$ defines a face $\mathcal{F}$ of the polyhedral cone $\mathcal{C}$. As in the Gaussian setting~\cite[Corollary 2.4]{LUZ}, the $\mtp$ MLE $\hat J$ is the MLE of the quadratic exponential family without the $\mtp$ constraint restricted to the face $\mathcal{F}$.  This is stated formally in Corollary~\ref{cor_polyhedral} and illustrated in Example~\ref{ex:mouss.alg} below.
\begin{cor}
	\label{cor_polyhedral}
	Let $\hat{J}$ denote the MLE in a quadratic exponential family under $\mtp$. Let $\mathcal{F}=\{(i,j)\in V\times V \mid \hat{J}_{ij}=0\}$. Then $\hat{J}$ equals the maximum likelihood estimate in the quadratic exponential family without the $\mtp$ constraint under the linear constraints $J_{ij}=0$ for all $(i,j)\in\mathcal{F}$.
\end{cor}
\begin{proof}This follows since the unique MLE in this quadratic exponential family is given by the equations (ii) and (iii) in Corollary~\ref{cor_MLE_MTP2} above. 
\end{proof}
In \cite{LUZ} it was shown that the MLE existed in the Gaussian case if and only if the empirical covariance matrix satisfied
$S_{ij}< \sqrt{S_{ii}S_{jj}}$ by constructing an ultrametric matrix $Z$ from $S$ that was both primary and dually feasible. The argument used in \cite{LUZ} does not apply here as the primary feasibility of $Z$ is not always guaranteed. Indeed, we shall see that the condition is necessary here but not sufficient; see Theorem~\ref{thm:fullMLE} and Corollary~\ref{cor:fullMLE} below.   The situation in a general exponential family can be quite different from the Gaussian case as shown in the following example.
\begin{ex}
	The auto-Poisson family considered in \cite[Section 4.2.4]{besag1974spatial} is a quadratic exponential family with product base. It consists of distributions of the form
	$$
	p(x; h,J)\propto \exp\left(\sum_{i=1}^d\left(h_ix_i-\log(x_i!)\right)+x^T J x/2\right)\qquad x\in \{0,1,2,\ldots\}^d.
	$$
	The right-hand side sums to a finite number if and only if $J_{ij}\leq 0$ for all $i,j$. The subset of $\mtp$ distributions within this family is then given by the product of independent Poisson distributions, that is, $J_{ij}=0$ for all $i,j$. 
Of course, for a finite state-space, no such problem occurs. 
\end{ex}

\section{Totally positive binary distributions}
\label{sec:binary}

For the remainder of this paper, we focus on binary distributions, i.e., distributions over the sample space $\cX=\{-1,1\}^d$. To simplify notation we often use the following bijection between $\cX$ and the set $\mathbf B_d$ of all subsets of $\{1,\ldots,d\}$, namely an element $x\in \cX$ maps to the subset of all $i\in \{1,\ldots,d\}$ for which $x_i=1$. For example, in the case $d=3$ the point $x=(1,1,-1)$ maps to the subset $\{1,2\}$ and $(-1,-1,-1)$ to the empty set. Note that $\cX$ and $\mathbf B_d$ are also isomorphic as lattices because the min-max operators $\wedge$, $\vee$ on $\cX$ correspond to the set operations $\cap$, $\cup$ in $\mathbf B_d$.  

Building on the results from Section~\ref{sec:expfamilies}, in the following we provide conditions for existence of the MLE in $\mtp$ binary exponential families. In particular, we study the KKT conditions for this setting and develop conditions for existence of the MLE in the special case of binary distributions that factorize according to a graph (such as Ising models) and symmetric binary distributions where $p(x)=p(-x)$ (such as Ising models with no external field). Ising models will be discussed in detail in Section~\ref{sec:ising}.

\subsection{Binary distributions as exponential families}

We now recall the representation of strictly positive binary distributions as an exponential family.  Define $\lambda(x):= \log p(x)$ for $x\in \cX= \{-1,1\}$. To write the exponential representation of this family of distributions we consider the space $\R^{\cX}$ of dimension $2^{d}$ equipped with the inner product 
$$
\langle \theta,\sigma\rangle \;\;:=\;\;\sum_{x\in \cX}\theta(x)\sigma(x).
$$
For $x\in \cX$, define a vector $T(x)\in \{0,1\}^{\cX}$ such that $T(x)_y=1$ if $x=y$ and it is zero otherwise. The set of binary distributions forms a regular exponential family which is minimally represented with canonical parameters  $\theta(x)=\lambda(x)-\lambda(-\mathbf 1)$ for $x\neq \mathbf -\mathbf 1$. Denote by $\theta$ the vector of all $\theta(x)$ for $x\in \cX$ and observe that $\theta(-\mathbf 1)=0$. Then
$$
p(x)=\exp(\langle \theta,T(x)\rangle - A(\theta)),
$$
where $A(\theta)=\log[\langle \mathbf 1,\exp(\theta)\rangle]$. The space of canonical parameters is simply the $2^{d}-1$ dimensional real vector space $\R^{\cX'}$ where $\cX'=\cX \setminus \{\mathbf{-1}\}$. The interior of the convex support of the sufficient statistics is given by the set
$$
\mathcal S=\left\{p \in \R^{\cX'}:\, p(x)>0\mbox{ for all } x\in \cX' \mbox{ and } \sum_{x\in \cX'}p(x) <1\right\},
$$which we identify with
the interior of the probability simplex, namely
$$
\mathcal{S}_1=\left\{p:\, p(x)>0\mbox{ for all }x\in \cX\mbox{ and } \sum_{x\in \cX}p(x) =1\right\}.
$$

\begin{rem}
The constraints on the space of canonical parameters $\cK$ defining binary $\mtp$ distributions are 
\begin{equation}\label{eq:MTP_binary}
\theta(x\wedge y)+\theta(x\vee y)-\theta(x)-\theta(y)\;\geq\;0
\end{equation}
for all elementary pairs $x,y\in \cX$. We recall that a pair $x,y$ is elementary if there exist a subset $A\subset V$ and $i,j\in V\setminus A$ such that $x$ corresponds to $A\cup \{i\}$ and $y$ corresponds to $A\cup \{j\}$. The number of such pairs is ${d \choose 2}2^{d-2}$. Another way to phrase  \eqref{eq:MTP_binary} is that $\theta$ is a supermodular set-function that satisfies the normalizing condition $\theta(-\mathbf 1)=0$; c.f.~\cite{bach2011learning}. 
\end{rem}
\subsection{KKT conditions and conditional independence}
\label{subsec_KKT}

In this section we study how the KKT conditions of Theorem~\ref{thm_mtp_exp_family} induce sparsity in the general binary setting, in the form of context-specific conditional independence constraints. 
To do this, we introduce some notation. Following Studen\'{y}~\cite{stu05}, we call the elements in $\mathbb Z^\cX$ \emph{imsets}. An important example of an imset is $T(x)\in \{0,1\}^{\cX}$ defined earlier. The imset
$$
u_{x,y}\;:=\;T(x\wedge y)+T(x\vee y)-T(x)-T(y)
$$
is called a \emph{semi-elementary imset}. If $x,y$ form an elementary pair then $u_{x,y}$ is called an \emph{elementary imset}. If this pair is associated to sets $A\cup \{i\}$ and $A\cup \{j\}$ we write $u_{i,j|A}$. With a slight abuse of notation we denote the class of all elementary imsets by $\mathcal E$.

Primal feasibility in Theorem~\ref{thm_mtp_exp_family} requires that $\hat \theta$ satisfies \eqref{eq:MTP_binary}, i.e.,
\begin{equation}\label{eq:MTP_binaryinner} 
	\langle \hat \theta,v\rangle\;\geq 0\qquad\mbox{for all }v\in \mathcal E.
\end{equation} 
The dual cone $\cC^\vee$ is the cone in $\R^\cX$ generated by all elementary imsets. 
Dual feasibility in Theorem~\ref{thm_mtp_exp_family} says that $\hat \sigma(x)>0$ for all $x\in \cX$ and 
\begin{equation}\label{eq:dualstud}
\hat \sigma-\bar T\;=\;\sum_{v\in \mathcal E}c_v v\qquad\mbox{where }c_v\geq 0.	
\end{equation}
Although every element in $\cC^\vee$ is a non-negative combination of elementary imsets, such a combination is typically not unique. For example, 
$$	u_{1,2|3}+u_{1,3|\emptyset}\;=\;u_{1,3|2}+u_{1,2|\emptyset}.$$
In particular, the coefficients $c_v$ above are not uniquely defined. But independent of the choice of these coefficients, the complementary slackness condition is equivalent to 
$$
\langle\hat\theta,\hat \sigma-\bar T\rangle\;=\;\sum_{v\in \mathcal E} c_v \langle \hat\theta,v\rangle\;=\; 0.
$$
By \eqref{eq:MTP_binaryinner}, this holds if and only if 
\begin{equation}\label{eq:binaryslackness}
c_v \langle \hat\theta,v\rangle\ \;=\; 0\qquad\mbox{for all }v\in \mathcal E.	
\end{equation}
We conclude that $\langle \hat\theta,v\rangle=0$ for every $v$ that appears in  a nonnegative linear combination of the form (\ref{eq:dualstud}). 
Therefore we obtain the following result. 
\begin{prop}Each equality in \eqref{eq:binaryslackness} corresponds to a context specific conditional independence statement where two variables are independent conditioned on a particular value of the remaining variables, as represented by an elementary imset.
\end{prop} 
\begin{proof}
Each inequality for a given elementary imset in \eqref{eq:MTP_binary} can be interpreted as a sign condition on a specific conditional correlation $${\rm cov}(X_i,X_j\cd X_{V\setminus \{i,j\}}=x)\geq 0,$$ corresponding to an elementary imset.\end{proof} Note that when $d=3$ there are six such constraints and these play an important role in the boundary decomposition of the latent class model \cite{allman2017maximum}. To see how they appear in the description of a general binary latent class model see \cite{ARSZ}. 
In the following example, we show how this characterization of complementary slackness can be used to compute the MLE.

\begin{ex}\label{ex:3}
	Let $d=3$ and consider the sample represented by the  diagram to the left in the following figure, where we again made use of the bijection between  $\{-1,1\}^3$ and the set of all subsets of $\{1,2,3\}$.

$$			\bar T\;=\;\frac{1}{13}\cdot\begin{tikzpicture}[scale=0.65,baseline=3ex]
\tikzstyle{every node}=[draw,shape=circle split, inner sep=0pt,minimum size =1.1cm];
  \node (max) at (0,4) { {${}^{\{1,2,3\}}$}\nodepart{lower} \begin{minipage}[c]{0.3cm}\vspace{3pt}$1$\end{minipage}};
  \node (a) at (-2,2) {{${}^{\{1,2\}}$}\nodepart{lower} \begin{minipage}[c]{0.3cm}\vspace{3pt}$2$\end{minipage}};
  \node (b) at (0,2) {{${}^{\{1,3\}}$}\nodepart{lower} \begin{minipage}[c]{0.3cm}\vspace{3pt}$0$\end{minipage}};
  \node (c) at (2,2) {{${}^{\{2,3\}}$}\nodepart{lower} \begin{minipage}[c]{0.3cm}\vspace{3pt}$4$\end{minipage}};
  \node (d) at (-2,0) {{${}^{\{1\}}$}\nodepart{lower} \begin{minipage}[c]{0.3cm}\vspace{3pt}$1$\end{minipage}};
  \node (e) at (0,0) {{${}^{\{2\}}$}\nodepart{lower} \begin{minipage}[c]{0.3cm}\vspace{3pt}$0$\end{minipage}};
  \node (f) at (2,0) {{${}^{\{3\}}$}\nodepart{lower} \begin{minipage}[c]{0.3cm}\vspace{3pt}$3$\end{minipage}};
  \node (min) at (0,-2) {{${}^\emptyset$}\nodepart{lower} \begin{minipage}[c]{0.2cm}\vspace{3pt}$2$\end{minipage}};
  \draw (min) -- (d) -- (a) -- (max) -- (b) -- (f)
  (e) -- (min) -- (f) -- (c) -- (max)
  (d) -- (b);
  \draw[preaction={draw=white, -,line width=6pt}] (a) -- (e) -- (c);
\end{tikzpicture}\qquad \hat \sigma\;=\;\frac{1}{182}\cdot\begin{tikzpicture}[scale=0.65,baseline=3ex]
\tikzstyle{every node}=[draw,shape=circle split, inner sep=0pt,minimum size =1.1cm];
  \node (max) at (0,4) { {${}^{\{1,2,3\}}$}\nodepart{lower} \begin{minipage}[c]{0.3cm}\vspace{3pt}$30$\end{minipage}};
  \node (a) at (-2,2) {{${}^{\{1,2\}}$}\nodepart{lower} \begin{minipage}[c]{0.3cm}\vspace{3pt}$12$\end{minipage}};
  \node (b) at (0,2) {{${}^{\{1,3\}}$}\nodepart{lower} \begin{minipage}[c]{0.3cm}\vspace{3pt}$7$\end{minipage}};
  \node (c) at (2,2) {{${}^{\{2,3\}}$}\nodepart{lower} \begin{minipage}[c]{0.3cm}\vspace{3pt}$40$\end{minipage}};
  \node (d) at (-2,0) {{${}^{\{1\}}$}\nodepart{lower} \begin{minipage}[c]{0.3cm}\vspace{3pt}$7$\end{minipage}};
  \node (e) at (0,0) {{${}^{\{2\}}$}\nodepart{lower} \begin{minipage}[c]{0.3cm}\vspace{3pt}$16$\end{minipage}};
  \node (f) at (2,0) {{${}^{\{3\}}$}\nodepart{lower} \begin{minipage}[c]{0.3cm}\vspace{3pt}$35$\end{minipage}};
  \node (min) at (0,-2) {{${}^\emptyset$}\nodepart{lower} \begin{minipage}[c]{0.2cm}\vspace{3pt}$35$\end{minipage}};
  \draw (min) -- (d) -- (a) -- (max) -- (b) -- (f)
  (e) -- (min) -- (f) -- (c) -- (max)
  (d) -- (b);
  \draw[preaction={draw=white, -,line width=6pt}] (a) -- (e) -- (c);
\end{tikzpicture}.
$$

We claim that $\hat\sigma$ represented by the diagram on the right corresponds to the MLE. First we check that $\hat\sigma$ is indeed $\mtp$ by checking that $\hat \sigma(x\vee y)\hat\sigma(x\wedge x)-\hat\sigma(x)\hat\sigma(y)\geq 0$ for all six elementary pairs $x,y$. Up to the normalizing constant $182$, these are
$$
\begin{array}{lcclc}
	\{1\}, \{2\}:& 12\cdot 35-7\cdot 16>0& \qquad &\{1,3\}, \{2,3\}:& 30\cdot 35-7\cdot 40>0 \\
	\{1\}, \{3\}:& \mathbf{7\cdot 35-7\cdot 35=0}& \qquad &\{1,2\}, \{2,3\}:& \mathbf{30\cdot 16-12\cdot 40=0} \\
	\{2\}, \{3\}:& 40\cdot 35-16\cdot 35>0& \qquad &\{1,2\}, \{1,3\}:& 	30\cdot 7-12\cdot 7>0 \\	
\end{array}.
$$
This proves primal feasibility in Theorem~\ref{thm_mtp_exp_family}. Dual feasibility is verified by the following diagram.
$$			\hat\sigma-\bar T\;=\;\frac{16}{182}\cdot \begin{tikzpicture}[scale=0.65,baseline=3ex]
\tikzstyle{every node}=[draw,shape=circle split, inner sep=0pt,minimum size =1.1cm];
  \node (max) at (0,4) { {${}^{\{1,2,3\}}$}\nodepart{lower} \begin{minipage}[c]{0.3cm}\vspace{3pt}$1$\end{minipage}};
  \node (a) at (-2,2) {{${}^{\{1,2\}}$}\nodepart{lower} \begin{minipage}[c]{0.3cm}\vspace{3pt}$-1$\end{minipage}};
  \node (b) at (0,2) {{${}^{\{1,3\}}$}\nodepart{lower} \begin{minipage}[c]{0.3cm}\vspace{3pt}$0$\end{minipage}};
  \node (c) at (2,2) {{${}^{\{2,3\}}$}\nodepart{lower} \begin{minipage}[c]{0.3cm}\vspace{3pt}$-1$\end{minipage}};
  \node (d) at (-2,0) {{${}^{\{1\}}$}\nodepart{lower} \begin{minipage}[c]{0.3cm}\vspace{3pt}$0$\end{minipage}};
  \node (e) at (0,0) {{${}^{\{2\}}$}\nodepart{lower} \begin{minipage}[c]{0.3cm}\vspace{3pt}$1$\end{minipage}};
  \node (f) at (2,0) {{${}^{\{3\}}$}\nodepart{lower} \begin{minipage}[c]{0.3cm}\vspace{3pt}$0$\end{minipage}};
  \node (min) at (0,-2) {{${}^\emptyset$}\nodepart{lower} \begin{minipage}[c]{0.2cm}\vspace{3pt}$0$\end{minipage}};
  \draw (min) -- (d) -- (a) -- (max) -- (b) -- (f)
  (e) -- (min) -- (f) -- (c) -- (max)
  (d) -- (b);
  \draw[preaction={draw=white, -,line width=6pt}] (a) -- (e) -- (c);
\end{tikzpicture}\;+\;\frac{7}{182}\cdot\begin{tikzpicture}[scale=0.65,baseline=3ex]
\tikzstyle{every node}=[draw,shape=circle split, inner sep=0pt,minimum size =1.1cm];
  \node (max) at (0,4) { {${}^{\{1,2,3\}}$}\nodepart{lower} \begin{minipage}[c]{0.3cm}\vspace{3pt}$0$\end{minipage}};
  \node (a) at (-2,2) {{${}^{\{1,2\}}$}\nodepart{lower} \begin{minipage}[c]{0.3cm}\vspace{3pt}$0$\end{minipage}};
  \node (b) at (0,2) {{${}^{\{1,3\}}$}\nodepart{lower} \begin{minipage}[c]{0.3cm}\vspace{3pt}$1$\end{minipage}};
  \node (c) at (2,2) {{${}^{\{2,3\}}$}\nodepart{lower} \begin{minipage}[c]{0.3cm}\vspace{3pt}$0$\end{minipage}};
  \node (d) at (-2,0) {{${}^{\{1\}}$}\nodepart{lower} \begin{minipage}[c]{0.3cm}\vspace{3pt}$-1$\end{minipage}};
  \node (e) at (0,0) {{${}^{\{2\}}$}\nodepart{lower} \begin{minipage}[c]{0.3cm}\vspace{3pt}$0$\end{minipage}};
  \node (f) at (2,0) {{${}^{\{3\}}$}\nodepart{lower} \begin{minipage}[c]{0.3cm}\vspace{3pt}$-1$\end{minipage}};
  \node (min) at (0,-2) {{${}^\emptyset$}\nodepart{lower} \begin{minipage}[c]{0.2cm}\vspace{3pt}$1$\end{minipage}};
  \draw (min) -- (d) -- (a) -- (max) -- (b) -- (f)
  (e) -- (min) -- (f) -- (c) -- (max)
  (d) -- (b);
  \draw[preaction={draw=white, -,line width=6pt}] (a) -- (e) -- (c);
\end{tikzpicture},
$$
In other words, $\hat\sigma-\bar T= \frac{16}{182}\cdot u_{1,3|2}+ \frac{7}{182}\cdot u_{1,3|\emptyset}\in\cC^\vee$. Complementary slackness follows by direct calculations. Note that the two nonzero generators in the decomposition of $\hat \sigma-\bar T$ correspond precisely to the $\mtp$ inequalities for $\hat \sigma$ that hold as equalities. These equalities correspond to the conditional independence statement $1\cip 3\cd 2$. \hfill\qed
\end{ex}

\subsection{Existence of the MLE}
\label{subsec_ex_MLE}
In this section we shall discuss problems associated with existence of the MLE for binary $\mtp$ distributions, the main result being Theorem~\ref{thm:fullMLE} which gives a simple necessary and sufficient condition for existence.

\subsubsection{Existence in the extended family}
To derive simple conditions for existence of the MLE within the exponential family of strictly positive binary distributions that are $\mtp$, we consider estimation in the extended family where the strict positivity condition is relaxed and existence therefore guaranteed. 

 Let $\P(\cX)$ denote the set of all probability distributions over $\cX$ and $\pto$ the set of all totally positive binary distributions, i.e.,
\[\pto \;=\; \{p\in \mathbb{P}(\mathcal X)\mid \forall x,y\in \mathcal{X}: p(x\vee y)p(x\wedge y)\geq p(x)p(y)\}.\]
We note that $\pto$ is compact and \emph{geometrically convex}, i.e.,
\[p_1,p_2\in \pto\implies c^{-1}\sqrt{p_1p_2}\in \pto\]
where 
\[c:= \sum_{x\in \spa} \sqrt{p_1(x)p_2(x)}\leq 1\]
and $c<1$ unless $p_1=p_2$ by the Cauchy--Schwarz inequality.

For a lattice $L$ we say that a subset $L'$ of $L$ forms a \emph{sublattice} of $L$ if for any two $x,y\in L$ it holds that $x\wedge y\in L'$ and $x\vee y\in L'$. 
Note that for any $p\in\pto$ its support ${\rm supp}(p)=\{x:p(x)>0\}$ is always a sublattice of $\spa$, since 
\[p(x)>0,\;  p(y)>0\implies p(x\vee y)p(x\wedge y)\geq p(x)p(y)>0.\]
Consider a sample $U=\{x^1,\ldots, x^n\}$ with likelihood function
\[L(p)=\prod_{i=1}^np(x^i)\]
and let $\mathcal{L}(U)$ be the the smallest sublattice of $\spa$ containing the sample $U$. We now show that the support of the MLE is given by $\mathcal{L}(U)$.

\begin{thm}\label{th:MLEbin}The likelihood function attains its maximum over $\pto$ in a unique point $\hat p$. Furthermore, it holds that   ${\rm supp}(\hat p)=\mathcal{L}(U)$.
\end{thm}
\begin{proof}Continuity of the likelihood function together with compactness of $\pto$ ensures that the maximum is attained. To prove uniqueness, suppose for contradiction that $\hat p_1\neq \hat p_2$ both maximize $L$. Then
	\[L(c^{-1}\sqrt{\hat p_1\hat p_2})=c^{-n}\sqrt{L(\hat p_1)L(\hat p_2)}>L(\hat p_i)\]
	contradicting that $\hat p_i$ were maximizers. 
	
	Finally, note that $U\subseteq {\rm supp}( \hat p)$ and hence $\mathcal{L}(U)\subseteq {\rm supp} (\hat p)$. We show $\mathcal{L}(U)\supseteq {\rm supp} (\hat p)$ by contradiction. Suppose $\mathcal{L}(U)\subsetneq {\rm supp} (\hat p)$, then we can construct $\tilde{p}\in\pto$ such that $L(\tilde p)> L(\hat p)$, which contradicts the fact that $\hat{p}$ is the MLE; namely, let $\tilde{p}$ be $\hat{p}$ projected onto $\mathcal{L}(U)$ and rescaled to be a probability mass function, i.e.\ $\tilde p(x)\propto p(x)\mathbf{1}_{\mathcal{L}(U)}$. Then $\tilde p\in \pto$ and $L(\tilde p)> L(\hat p)$, which concludes the proof.
\end{proof}
\subsubsection{Existence of MLE in the binary exponential family}
The MLE exists in the binary exponential family if and only if the estimator $\hat p$ in the extended family $\mathbb P(\cX)$ has full support. Thus as a consequence of Theorem~\ref{th:MLEbin} we obtain the following result, where $U_{ij}=\{x^1_{ij},\ldots,x^n_{ij}\}$ denotes the marginal sample induced on the pair $ij,  i\neq j$ . 
\begin{thm}\label{thm:fullMLE}
The MLE exists within the space of totally positive canonical parameters $\cK_2$ (c.f. Definition~\ref{def:K2})  if and only if $\mathcal{L}(U)=\cX$. Furthermore, $\mathcal{L}(U)=\cX$ if and only if every pair-marginal sample  $U_{ij}$ for  $i,j\in V=\{1,\ldots,d\}$ has both  
 of $(1,-1)$  and $(-1,1)$ represented.
 \end{thm}
 \begin{proof}
As mentioned, the MLE exists in the binary exponential family if and only if the estimator $\hat p$ in the extended family $\mathbb P(\cX)$ has full support. Thus, as a consequence of Theorem~\ref{th:MLEbin}, the $\mtp$ MLE exists if and only $\mathcal{L}(U)=\cX$. 
 	
 For the second statement we first prove the backward direction	using the identification between $\cX$ and subsets of $V$. Suppose every pair-marginal $U_{ij}$ for  $i,j\in V$ has both  
 of $(1,-1)$  and $(-1,1)$ represented. This means that for every $i$ there is a set $x_{ij}\in U$ with $i\in x_{ij}$ and $j\not \in  x_{ij}$. But then 
\[ \{i\} = \bigcap_{j\in V\setminus i}x_{ij} \in \mathcal{L}(U) \text{ for all $i$.}\]
Since the set of all singletons $\{i\}$ for  $i\in V$ generates the full lattice $\cX$, we obtain $\mathcal{L}(U)=\cX$ as desired.
 
We prove the forward direction by proving its contrapositive. Suppose there is a pair $ij$ such that all sets $x\in U$ have the property that 
 \begin{equation}\label{eq:degen} i\in x\implies j\in x.
 \end{equation}
 The set of subsets $y$ satisfying (\ref{eq:degen}) form a proper sublattice $\mathcal{L}'\subset \cX$. Since $\mathcal{L}(U)\subseteq \mathcal{L}'$ we obtain that $\mathcal{L}(U)\neq \cX$, which completes the proof.
 \end{proof}

	Theorem 1 in \cite{slawski2015estimation} states that the MLE in the $\mtp$ Gaussian distribution exists if and only if all sample correlations are strictly less than one. Theorem~\ref{thm:fullMLE} yields the analogous result for  binary distributions. Indeed we have the following.
	\begin{cor}\label{cor:fullMLE}If the MLE exists within $\cK_2$, then the empirical covariance matrix satisfies $S_{ij}<\sqrt{S_{ii}S_{jj}}$ for all $i\neq j$.
	\end{cor}
	\begin{proof}The empirical correlation matrix $R$ has $|R_{ij}|=1$ if and only if it holds for all $x\in U$ in the sample that $x_j=ax_i+b$.  If both configurations $(1,-1)$ and $(-1,1)$ are represented in $U$, this would imply $b-a=1$ and $b+a=-1$ whereby $b=0$, $a=-1$ and thus $R_{ij}=-1$ implying $S_{ij}<\sqrt{S_{ii}S_{jj}}$.  
	\end{proof}
	Note that the converse is not true.  If for two variables the sample is $U=\{(-1,-1), (1,-1), (1,1)\}$, then the MLE does not exist according to Theorem~\ref{thm:fullMLE}, but we have $S_{11}=S_{22}=8/9$ and $S_{12}= 4/9$; so the empirical correlation is equal to $1/2$.

As another example, consider the case $d=3$. Then the vectors $(1,-1,-1)$, $(-1,1,-1)$, $(-1,-1,1)$ generate all of $\{-1,1\}^3$ and hence every sample supported on these three points will admit a unique MLE under the $\mtp$ constraint. This set is minimal in the sense that it cannot be reduced; none of its subsets generates $\cX$. There are also minimal generating subsets of size four, e.g.~$(1,1,-1)$, $(1,-1,-1)$, $(-1,-1,1)$, $(-1,1,1)$. For general $d$, a minimal generating set of $\{-1,1\}^d$ is of order $\mathcal O(d)$ and there always exists a minimal generating set of size exactly $d$. Hence for binary $\mtp$ distributions $d$ samples can be sufficient for existence of the MLE. This is in sharp contrast with unrestricted binary exponential families, where the MLE exists only if \emph{all} $2^d$ states are observed at least once.

While the MLE in Example~\ref{ex:3} could be computed by hand, calculations get intractable rather quickly. 
The following example 
is sufficiently complicated that it cannot easily be calculated by hand, but still simple enough so that numerical optimization using the algorithm developed in \cite{forcina2000} yields the provably exact optimum. 
\begin{ex}\label{ex:moussouris}
	Moussouris \cite{moussouris1974gibbs} provided a now classical example of a distribution $q$ that is globally Markov to its dependence graph but does not factorize; c.f. \cite[Example~3.10]{lau96}. The distribution in this example is uniformly supported on eight points 
$$
\begin{array}{llll}
	(-1,-1,-1,-1) & (1,-1,-1,-1) & (1,1,-1,-1) & (1,1,1,-1)\\ 
		(-1,-1,-1,1) & (-1,-1,1,1) & (-1,1,1,1) & (1,1,1,1).		
\end{array}
$$
This distribution is globally Markov with respect to the 4-cycle in Figure~\ref{fig:4chain} (left),
\begin{figure}[!t]
	\begin{center}
		\begin{tikzpicture}[scale=.6]
		\draw[fill=black] (0,0) circle (6pt);
		\draw[fill=black] (2,0) circle (6pt);
		\draw[fill=black] (0,2) circle (6pt);
		\draw[fill=black] (2,2) circle (6pt);
		\node at (-0.4,2.4) {1};
		\node at (2.4,2.4) {2};
		\node at (2.4,-0.4) {3};
		\node at (-0.4,-0.4) {4};
		\draw[thick] (0,0) -- (2,0) --  (2,2) -- (0,2) -- (0,0);
		\end{tikzpicture}	\qquad\qquad
		\begin{tikzpicture}[scale=.6]
		\draw[fill=black] (0,0) circle (6pt);
		\draw[fill=black] (2,0) circle (6pt);
		\draw[fill=black] (0,2) circle (6pt);
		\draw[fill=black] (2,2) circle (6pt);
		\node at (-0.4,2.4) {1};
		\node at (2.4,2.4) {2};
		\node at (2.4,-0.4) {3};
		\node at (-0.4,-0.4) {4};
		\draw[thick] (0,0) -- (2,0) --  (2,2) -- (0,2);
		\end{tikzpicture}	
	\end{center}
\vspace{-0.4cm}
	\caption{A cycle (left) and a chain (right) with four vertices.}\label{fig:4chain}
\end{figure}
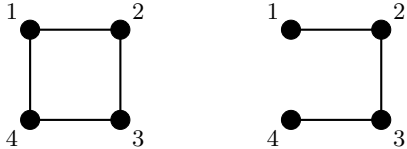
and we shall consider these eight points as constituting a sample of size eight. 
The MLE for this graphical model as an exponential family does not exist. Note that the sample distribution is not $\mtp$, since for example the inequality $$p(1,-1,-1,1) p(-1,-1,-1,-1)\geq p(1,-1,-1,-1)p(-1,1,-1,-1)$$ does \emph{not} hold. On the other hand, since the conditions of Theorem \ref{thm:fullMLE} are satisfied, the MLE $\hat{\sigma}$ under $\mtp$ exists. It is represented by the following diagram, where the highlighted nodes correspond to the eight points supported by the sample.
$$		\hat \sigma\;=\;\frac{1}{128}\cdot\begin{tikzpicture}[scale=0.7,baseline=-.5ex]
\tikzstyle{every node}=[draw,shape=circle split, inner sep=0pt,minimum size =1.2cm];
  \node[fill=yellow] (max) at (0,4.5) { {${}^{\{1,2,3,4\}}$}\nodepart{lower} \begin{minipage}[c]{0.3cm}\vspace{3pt}$27$\end{minipage}};
  \node[fill=yellow] (123) at (-3,2.5) {{${}^{\{1,2,3\}}$}\nodepart{lower} \begin{minipage}[c]{0.3cm}\vspace{3pt}$9$\end{minipage}};
  \node (124) at (-1,2.5) {{${}^{\{1,2,4\}}$}\nodepart{lower} \begin{minipage}[c]{0.3cm}\vspace{3pt}$3$\end{minipage}};
  \node (134) at (1,2.5) {{${}^{\{1,3,4\}}$}\nodepart{lower} \begin{minipage}[c]{0.3cm}\vspace{3pt}$3$\end{minipage}};
  \node[fill=yellow] (234) at (3,2.5) {{${}^{\{2,3,4\}}$}\nodepart{lower} \begin{minipage}[c]{0.3cm}\vspace{3pt}$9$\end{minipage}};
  \node[fill=yellow] (12) at (-5,0) {{${}^{\{1,2\}}$}\nodepart{lower} \begin{minipage}[c]{0.3cm}\vspace{3pt}$9$\end{minipage}};
  \node (13) at (-3,0) {{${}^{\{1,3\}}$}\nodepart{lower} \begin{minipage}[c]{0.3cm}\vspace{3pt}$1$\end{minipage}};
  \node (14) at (-1,0) {{${}^{\{1,4\}}$}\nodepart{lower} \begin{minipage}[c]{0.3cm}\vspace{3pt}$3$\end{minipage}};
  \node (23) at (1,0) {{${}^{\{2,3\}}$}\nodepart{lower} \begin{minipage}[c]{0.3cm}\vspace{3pt}$3$\end{minipage}};
  \node (24) at (3,0) {{${}^{\{2,4\}}$}\nodepart{lower} \begin{minipage}[c]{0.3cm}\vspace{3pt}$1$\end{minipage}};
  \node[fill=yellow] (34) at (5,0) {{${}^{\{3,4\}}$}\nodepart{lower} \begin{minipage}[c]{0.3cm}\vspace{3pt}$9$\end{minipage}};
  \node[fill=yellow] (1) at (-3,-2.5) {{${}^{\{1\}}$}\nodepart{lower} \begin{minipage}[c]{0.3cm}\vspace{3pt}$9$\end{minipage}};
  \node (2) at (-1,-2.5) {{${}^{\{2\}}$}\nodepart{lower} \begin{minipage}[c]{0.3cm}\vspace{3pt}$3$\end{minipage}};
  \node (3) at (1,-2.5) {{${}^{\{3\}}$}\nodepart{lower} \begin{minipage}[c]{0.3cm}\vspace{3pt}$3$\end{minipage}};
  \node[fill=yellow] (4) at (3,-2.5) {{${}^{\{4\}}$}\nodepart{lower} \begin{minipage}[c]{0.3cm}\vspace{3pt}$9$\end{minipage}};
  \node[fill=yellow] (min) at (0,-4.5) {{${}^\emptyset$}\nodepart{lower} \begin{minipage}[c]{0.3cm}\vspace{3pt}$27$\end{minipage}};
  \draw (min) -- (1) -- (12) -- (123) -- (max) -- (124) -- (14) -- (4) -- (min) -- (1) -- (14) -- (134) -- (max);
  \draw (3) -- (min) -- (1) -- (13) -- (123) -- (max) -- (234) -- (34) -- (4) -- (24) -- (234);
  \draw[preaction={draw=white, -,line width=6pt}] (min) -- (2) -- (12) -- (124) -- (24) -- (2) -- (23) -- (234);
  \draw[preaction={draw=white, -,line width=6pt}] (3) -- (13) -- (134) -- (34) -- (3) -- (23) -- (123);
\end{tikzpicture},
$$
Primal feasibility of $\hat \sigma$ is verified by the following inequalities, one for each of the $24$ elementary pairs (labeled by sets $\{i\}\cup A$ and $\{j\}\cup A$). Up to the normalizing constant $128$, these are: 
$$
{\small\begin{array}{lcclc}
	\{1\},\{2\}:& & 9\cdot 27-9\cdot 3>0$\qquad$ & \{1,3\},\{2,3\}: & 9\cdot 3-1\cdot 3>0\\
		\{1,4\},\{2,4\}:& & 3\cdot 9-3\cdot 1>0$\qquad$ & \{1,3,4\},\{2,3,4\}: & 27\cdot 9-3\cdot 9>0\\
	\{1\},\{3\}:& & \mathbf{1\cdot 27-9\cdot 3=0}$\qquad$ & \{1,2\},\{2,3\}: & \mathbf{9\cdot 3-9\cdot 3=0}\\
		\{1,4\},\{3,4\}:& & \mathbf{3\cdot 9-3\cdot 9=0}$\qquad$ & \{1,2,4\},\{2,3,4\}: & \mathbf{27\cdot 1-3\cdot 9=0}\\
			\{1\},\{4\}:& & \mathbf{3\cdot 27-9\cdot 9=0}$\qquad$ & \{1,2\},\{2,4\}: & \mathbf{3\cdot 3-9\cdot 1=0}\\
		\{1,3\},\{3,4\}:& & \mathbf{3\cdot 3-1\cdot 9=0}$\qquad$ & \{1,2,3\},\{2,3,4\}: & \mathbf{27\cdot 3-9\cdot 9=0}\\
		\{2\},\{3\}:& & 3\cdot 27-3\cdot 3>0$\qquad$ & \{1,2\},\{1,3\}: & 9\cdot 9-9\cdot 1>0\\
		\{2,4\},\{3,4\}:& & 9\cdot 9-1\cdot 9>0$\qquad$ & \{1,2,4\},\{1,3,4\}: & 27\cdot 3-3\cdot 3>0\\
		\{2\},\{4\}:& & \mathbf{1\cdot 27-3\cdot 9=0}$\qquad$ & \{1,2\},\{1,4\}: & \mathbf{3\cdot 9-9\cdot 3=0}\\
		\{2,3\},\{3,4\}:& & \mathbf{9\cdot 3-3\cdot 9=0}$\qquad$ & \{1,2,3\},\{1,3,4\}: & \mathbf{27\cdot 1-9\cdot 3=0}\\
			\{3\},\{4\}:& & 9\cdot 27-3\cdot 9>0$\qquad$ & \{1,3\},\{1,4\}: & 3\cdot 9-1\cdot 3>0\\
		\{2,3\},\{2,4\}:& & 9\cdot 3-3\cdot 1>0$\qquad$ & \{1,2,3\},\{1,2,4\}: & 27\cdot 9-9\cdot 3>0
		\end{array}}
$$
Quite surprisingly the MLE is therefore still globally Markov to the 4-cycle even though these constraints were not explicitly enforced. Moreover, $\hat\sigma$ satisfies an additional conditional independence relation, namely $1\indep 4|\{2,3\}$, and so it is Markov to the smaller graph in Figure~\ref{fig:4chain} (right).

There are many equivalent ways to write the vector $\hat\sigma-\bar T$. The most canonical is the one using all twelve elementary imsets allowed by the complementary slackness condition \eqref{eq:binaryslackness}, that is, the ones corresponding to boldfaced rows above:
\begin{eqnarray*}
	\hat\sigma-\bar T &=& \frac{3}{128} \cdot u_{1,3|\emptyset}+ \frac{1}{128}\cdot u_{1,3|2}+ \frac{1}{128}\cdot u_{1,3|4}+\frac{3}{128}\cdot u_{1,3|2,4}+\\
	&+&\frac{3}{128} \cdot u_{2,4|\emptyset}+\frac{1}{128} \cdot u_{2,4|1}+ \frac{1}{128}\cdot u_{2,4|3}+\frac{3}{128} \cdot u_{2,4|1,3}+\\
	&+& \frac{5}{128}\cdot u_{1,4|\emptyset}+\frac{5}{128}\cdot u_{1,4|2}+\frac{5}{128}\cdot u_{1,4|3}+\frac{5}{128}\cdot u_{1,4|2,3}.
\end{eqnarray*}
Each of the vectors $u_{i,j|A}$ above is a generator of $\cC^\vee$ and so $\hat \sigma-\bar T\in \cC^\vee$. \hfill\qed 

\end{ex}

\begin{rem}
	To show that $\hat\sigma-\bar T$ lies in $\cC^\vee$ it is enough to express it as a nonnegative combination of vectors $T(x\wedge y)+T(x\vee y)-T(x)-T(y)$ for \emph{arbitrary} pairs $x,y\in \cX$. This follows directly from \cite[Proposition 4.2]{stu05}.
\end{rem}

Note that in the above examples the MLEs correspond to models satisfying conditional independence statements. However, in general the MLE will satisfy a set of context specific conditional independence statements that may not lead to full conditional independences. In the following subsection, we consider binary $\mtp$ models that satisfy conditional independence relations given by a graphical model.

\subsection{Totally positive graphical models for binary variables}
\label{sec_models_graphs}

Given a graph $G=(V,E)$, let $\mathcal P_2(G)$ denote the set of distributions in $\mathcal P_2$  that lie in the completion of the exponential family (\cite[pp.\ 154-155]{barndorff:78}) for the graphical model over $G$, i.e.
\[\mathcal{P}_2(G) = \mathcal P_2 \cap M_E(G),\]
where $M_E(G)$ denotes the set of \emph{extended Markov} distributions (\cite[p. 40]{lau96}) obtained as limits of factorizing distributions; see also \cite{geiger:etal:06}. We note that $\mathcal P_2(G)$ is compact and geometrically convex (see e.g.\ \cite[p.\ 73]{lau96}); hence the MLE over $\mathcal P_2 (G)$ exists and is unique.

We first need a lemma to identify when binary $\mtp$ distributions $p\in\mathcal P_2$ have full support based on their marginals. These results are critical for this section in order to identify when the MLE of a binary distribution that is Markov over a graph has full support.

\begin{lem}\label{lem:bigconj}
	Let $p\in \mathcal P_2$ and let $x\in \cX$. Suppose $p_{ij}(x_i,x_j)>0$ for all pairs $i,j$ then $p(x)>0$. \end{lem}
\begin{proof}
	For every $i,j$ let $y^{(ij)}\in {\rm supp}(p)$ such that $y^{(ij)}_{ij}=(x_i,x_j)$. Let $A/B$ be the partition of $V$ such that $x_i=-1$ for $i\in A$ and $x_i=1$ on $B$. For each $i\in A$ define $z^{(i)}=\max_{j\in B}y^{(ij)}$. By construction $z^{(i)}_i=-1$ and $z^{(i)}_B=(1,\ldots,1)$. Moreover $z^{(i)}\in {\rm supp}(p)$ because ${\rm supp}(p)$ is a lattice. Since $x=\min_{i\in A} z^{(i)}$, $x\in {\rm supp}(p)$ again because the support of $p$ is a lattice.
\end{proof}

\begin{cor}\label{cor:pair}
	If $p\in \mathcal P_2$ then $p$ has full support $\cX$ if and only if each pair-margin $p_{ij}$ has full support. 
\end{cor}

The following result extends Theorem \ref{thm:fullMLE} to binary graphical models and relaxes the pair-marginal condition to be necessary only for pairs of neighbours in the graph $G$. As before $U_{ij}=\{x^1_{ij},\ldots,x^n_{ij}\}$ denotes the pair-marginal sample for the pair $ij$.

\begin{thm}\label{th:mainMLE}
If every pair marginal sample  $U_{ij}$ along edges $ ij\in E$ has both of $(1,-1)$ and $(-1,1)$ represented, then
the unique MLE $\hat p\in\mathcal P_2(G)$ has full support.	
\end{thm}

The proof makes use of the fact that the support of $\hat{p}$, denoted by $\Supp (\hat p)$, is a lattice since $\hat p \in \mathcal P_2$. In addition, since $\hat p\in M_E(G)$,
$\hat{p}$ also satisfies the global, local, and pairwise Markov properties w.r.t.\ $G$ (\cite[p. 42, (3.16)]{lau96}. 
In particular, the proof relies  on the following two lemmas. 
\begin{lem}\label{lem:edges}
	If the pair marginal sample  $U_{ij}$ has both of $(1,-1)$ and $(-1,1)$ represented for all $ij\in E$, then $\Supp(\hat p_{ij}) =\{-1,1\}^2$ for all $ij\in E$. 
\end{lem}
\begin{proof}
The $\mtp$ property is closed under taking marginals (see \cite{KarlinRinott80}). So  if $\hat p$ is $\mtp$, so are its marginals $\hat p_{ij}$. Thus $\Supp(\hat p_{ij})$ is a lattice containing $U_{ij}$. As a consequence, if $U_{ij}$ has both of $(1,-1)$ and $(-1,1)$ represented, then $\Supp(\hat p_{ij})= \{-1,1\}^2$, which completes the proof.
\end{proof}

Denoting by $\partial i$ the neighbors of node $i\in V$ in $G$, the following lemma will be needed for showing that $\Supp(\hat p_{ij})=\{-1,1\}^2$ for \emph{all} pairs $ij$ and not only the pairs $ij\in E$.

\begin{lem}\label{lem:support2}
Suppose that every  pair marginal sample $U_{ij}$ along edges $ij\in E$ has both of $(1,-1)$ and $(-1,1)$ represented. If $\,\hat p_{\partial i}(x_{\partial i})>0$ for some $x_{\partial i}$, then $\hat p_{i\cup \partial i}(x_{i\cup\partial i})>0$ for every $x_i$. 
\end{lem}
\begin{proof}
	Since $\hat p_{\partial i}(x_{\partial i})>0$, clearly $\hat p_{i\cup \partial i}(x_{i\cup\partial i})>0$ for some $x_i$, say $x_i=1$. We need to show that $\hat p_{i\cup \partial i}(y_{i\cup\partial i})>0$ also if $y_i=-1$ and $y_{\partial i}=x_{\partial i}$. Let $z_{i\cup \partial i}$ be such that $z_i=-1$ and $z_{\partial i}=(1,\ldots,1)$. Since $\hat p\in \mathcal P_2(G)$, its support is a lattice and the same applies to each margin of $\hat p$. Because 
	$$
	y_{i\cup \partial i}\;=\;x_{i\cup \partial i} \wedge z_{i\cup \partial i},
	$$  
	to show that $y_{i\cup \partial i}$ lies in the support of $\hat p_{i\cup \partial i}$ it is sufficient to show that this holds for $z_{i\cup \partial i}$. By the assertion, for each $j\in \partial i$ the edge-margin $U_{ij}$ has $(-1,1)$ represented. In particular,  there is a point $u^{(j)}\in \cX$ such that $u^{(j)}_{i}=-1$ and $u^{(j)}_j=1$. The support of $\hat p$ necessarily contains all elements in $U$ and hence $\hat p(u^{(j)})>0$ for all $j\in \partial i$. Let $u$ be the elementwise maximum of all $u^{(j)}$. This point lies in $\Supp (\hat p)$ because it forms a lattice. By construction, $u_{i\cup \partial i}=z_{i\cup \partial i}$, which proves that $z_{i\cup \partial i}$ (and hence also $y_{i\cup \partial i}$) lies in the support of $\hat p_{i\cup \partial i}$. The proof for the case where $x_i=-1$ is analogous.
\end{proof}

We are now ready to provide the proof of Theorem~\ref{th:mainMLE}.
\begin{proof}[Proof of Theorem~\ref{th:mainMLE}] From Corollary~\ref{cor:pair} it follows that $\hat p$ has full support if and only if the marginal support $\Supp(\hat p_{ij})$ is full for all $i,j\in V$. When $ij\in E$, this follows from Lemma~\ref{lem:edges}. Next, consider a pair $ij\notin E$.  
	Since $\hat p\in M_E(G)$, it satisfies the local Markov property with respect to $G$. Hence for any $x_i,x_j\in \{-1,1\}$, it holds that
\begin{eqnarray*}
	\hat p_{ij}(x_i,x_j)&=& \sum_{x_{\partial i\cup \partial j}}\hat p(x_i,x_j\cd x_{\partial i\cup \partial j})\hat p(x_{\partial i\cup \partial j})\\&=&\sum_{x_{\partial i\cup \partial j}} \hat p(x_i\cd x_{\partial i})\hat p(x_j\cd x_{\partial j})\hat p(x_{\partial i\cup \partial j}).
\end{eqnarray*}
	Since there is at least one $x_{\partial i\cup \partial j}$ in the support of $\hat p_{\partial i\cup \partial j}$, then by Lemma~\ref{lem:support2} both of $\hat p(x_i, x_{\partial i})$ and  $\hat p(x_j, x_{\partial j})$ are strictly positive and hence also the corresponding summand. It follows that $\hat p_{ij}(x_i,x_j)>0$, as desired. 
\end{proof}

Theorem~\ref{th:mainMLE} provides conditions for the existence of the MLE in the underlying exponential family, which we denote by $\mathcal K_2(G)$, consisting of all points in $\mathcal P_2(G)$ with full support.  

\begin{cor}\label{cor:bipartite}
If $G$ is bipartite, then the minimal sample size required for existence of the MLE is $n=2$. 
More generally, for arbitrary graphs the minimal sample size for existence of the MLE is of the order of the maximal clique size.
\end{cor}

 Hence the minimal sample size for existence of the MLE goes from $2^d$ for unrestricted binary distributions, to $d$ for $\mtp$ binary  distributions, to $\mathcal O(\textrm{maximal clique size})$ for $\mtp$ binary  distributions on graphs, including Ising models. In the following subsection, we consider a special class of binary distributions that contain as prominent examples Ising models without external field and show that the minimal sample size for existence of the MLE can be further reduced.

\subsection{Symmetric binary distributions}
\label{subsec_symmetric}

A distribution $p$ over $\cX=\{-1,1\}^d$ is \emph{symmetric} (or \emph{palindromic}) if $p(x)=p(-x)$ for all $x\in \cX$. Distributions of this form have been studied for example in \cite{marchetti:wermuth:16} and also appear in statistical physics in the context of spin models with no external field. If $X=(X_1,\ldots,X_d)$ has a symmetric distribution, then $\E X_i=0$ and ${\rm var}(X_i)=1$ for all $i=1,\ldots,d$. As a consequence, the covariance matrix and the correlation matrix of $X$ coincide. 
Note also that symmetry translates into linear constraints $\theta(x)=\theta(-x)$ for all $x\in \cX$ on the canonical parameters of the binary exponential family. Hence symmetric distributions with full support form themselves an exponential family. In the following, we characterize existence of the MLE for symmetric binary distributions.

Let as before $U=\{x^1,\ldots,x^n\}$ denote a random sample. Let $\cA(U)$ denote the smallest algebra generated by $U$, that is, the smallest subset of $\cX$ that contains $U$ and is closed under the lattice operations $\wedge,\vee$ and the complement $x\mapsto -x$.  For a family of distributions $\mathcal{P}$ we let $\mathcal P^{s}$ denote the set of symmetric distributions in $\mathcal{P}$ and  $U^s= U\cup -U$ be the symmetrized sample. 
\begin{prop}\label{prop:symcounts}If $\mathcal{P}$ is geometrically convex, then the MLE $\hat p_s$ under $\mathcal{P}^s$  based on a sample $U$  exists in $\mathcal{P}^s$ if and only if the MLE $\tilde p_s$ under $\mathcal{P}$ based on the symmetrized sample $U^s$ exists. In this case, it holds that $\hat p_s=\tilde p_s$. 
\end{prop}
\begin{proof}
Note that for any $p\in \mathcal P$, the likelihood function satisfies
\[ L(p; U^s) = \prod_{x\in \cX} p(x)^{n(x)+n(-x)}=  L(\check p; U^s), \]
where $\check p(x)=p(-x)$,
and  $n(x)= |\{i\in 1,\ldots, n: x_i=x\}|$ are the empirical counts in the sample $U$. Since $\mathcal{P}$ is geometrically convex, a maximizer $\tilde p_s$ of  $L(p; U^s)$ is unique; thus $\tilde p_s(x)=\tilde p_s(-x)$ and hence $\tilde p \in\mathcal{P}^s$. 
Note also that for any $p_s\in \mathcal{P}^s$ we have
\[L(p_s;U)^2  = L( p_s; U^s).\]
So any maximizer of $L(p_s;U)$ over $\mathcal {P}^s$ is also a maximizer of $L( p_s; U^s)$ and vice-versa.  Finally, the uniqueness implies that $\hat p_s=\tilde p_s$, as desired.
\end{proof}

By combining Proposition~\ref{prop:symcounts} with Theorem~\ref{thm:fullMLE} we obtain the following corollary on the existence of the MLE for symmetric binary distributions.
\begin{cor}\label{prop:fullMLEsymm}
The MLE for a symmetric binary exponential family exists if and only if $\mathcal{A}(U)=\cX$. Furthermore, $\mathcal{A}(U)=\cX$ if and only if for every pair $ij$ the event $\{X_i\neq X_j\}$ is represented in the sample.
\end{cor}

Finally, as a consequence we obtain the following corollary as an application of Theorem~\ref{th:mainMLE} to symmetric binary distributions on graphs defined as $\mathcal P_2^{s}(G):=\mathcal P_2(G)\cap \mathcal P_2^{s}$. 
\begin{cor}\label{cor_full_s}
If  the event $\{X_i\neq X_j\}$ is represented in every pair marginal sample $U_{ij}$, then the MLE $\hat p$ in the family $\mathcal P_2^s(G)$ has full support.	
\end{cor}

\begin{rem}We note again the remark to Theorem 1 in \cite{slawski2015estimation} which states that the MLE in an $\mtp$ Gaussian distribution exists if and only if all sample correlations are strictly less than one. Corollary~\ref{cor_full_s} implies that \emph{exactly the same is true for symmetric binary distributions.} Interestingly, while for (nontrivial, i.e., with at least one edge) Gaussian graphical models sample size equal to two is necessary and sufficient for existence of the MLE (with probability 1)~\cite{LUZ}, as a consequence of Corollary~\ref{cor:bipartite} and Corollary~\ref{cor_full_s}, the MLE for a symmetric binary distribution on a bipartite graph may have full support for sample size equal to one. 
\end{rem}

\section{Totally positive Ising models}\label{sec:ising}

In this section, we study maximum likelihood estimation in \emph{Ising models}, a special class of binary distributions that form a quadratic exponential family. An algorithm for calculating the MLE $\hat{p}$ for general binary $\mtp$ distributions was developed in \cite{forcina2000}. In Section~\ref{subsec:IPS}, we develop an algorithm analogous to \emph{iterative proportional scaling} (\emph{IPS}) for the special case of Ising models under $\mtp$. In addition, we discuss the special case of $\mtp$ Ising models with no external field, which forms a symmetric exponential family. Such distributions can be seen as a proxy to Gaussian distributions and in Section~\ref{subsec:noext} we discuss their similarities and differences.

Since Ising models form a quadratic exponential family, their probability mass function is of the form 
\begin{equation}\label{eq:ising}p(x;h,J)\;\;=\;\;\exp\left(h^T x+x^T J  x/2-A(h,J )\right),	\end{equation}
with $h\in \R^d$ and $J \in \S^d_0$, where $\S^d_0$ is the set of symmetric matrices in $\R^{d\times d}$ with $J_{ii}=0$ for all $i$, ensuring minimality of the representation; see (\ref{eq:quadratic}). 
We let $\mathcal{I}_2$ be the set of Ising models above that are also $\mtp$, i.e.\ where $J_{ij}\geq 0$ for all $i\neq j$.

Let $\theta=(h,J)$ denote the canonical parameters. We make the following two important observations regarding the canonical parameters. For any $i,j\in V$ let $A= V\setminus \{i,j\}$. Then the corresponding conditional log-odds ratios are all equal;  more precisely, denote by $x,y\in \cX$ any two points satisfying $x_A=y_A$, $x_i=y_j=1$, and $x_j=y_i=-1$, then
\begin{equation}\label{eq:logoddsIsing}
\log\left(\frac{p(x\vee y)p(x\wedge y)}{p(x)p(y)}\right)\;=\;4J_{ij}.	
\end{equation}
This is another way of confirming that an Ising model defined by $(h,J)$ is $\mtp$ if and only if $J\in \mathbb S_+^d\cap \mathbb{S}^d_0$; see Proposition~\ref{prop:mtpquadratic}. In addition, note that for any $x$ with $x_i=1$ and $y$ equal to $x$ up to the i'th coordinate, then
\begin{equation}\label{eq:ratioIsing}
\log\left(\frac{p(x)p(-y)}{p(-x)p(y)}\right)\;=\;4h_i.	
\end{equation}

The sufficient statistics based on the observations $x^{1},\ldots, x^n$ are the first and second order moments
\[(\bar x, M):=\frac{1}{n}\left(\sum_{i=1}^n{x^i}, \sum_{i=1}^n x^i (x^i)^T\right).\]
Strictly speaking we should ignore the diagonal elements of $M$, but since they are all deterministically equal to $1$, this does not matter for the following considerations. In addition, for a graphical Ising model on $G=(V,E)$ --- i.e.\ where we assume $J_{ij}=0$ unless $ij\in E$ --- the entries $M_{ij}$ for $ij\not\in E$ should be ignored. 
The associated \emph{mean value parameters} are 
\[(\mu,\Xi):= (\E_\theta X, \E_\theta X X^T).\]

\subsection{Existence of the MLE for totally positive Ising models}
 Theorem~\ref{th:mainMLE} can be specialized to the quadratic case, i.e.\ when also the Ising model is assumed. The condition for existence is here unchanged compared to the general Markov case.  For an undirected graph $G=(V,E)$ let $\mathcal{I}_2(G)$ be the family of totally positive Ising models that are Markov w.r.t.\ $G$, i.e.\ where $J_{ij}=0$ unless $ij\in E$. We then have

\begin{thm}\label{thm:mle_ising}
If every pair marginal sample  $U_{ij}$ along edges $ ij\in E$ has both of $(1,-1)$ and $(-1,1)$ represented, then
the MLE $\hat p\in \mathcal{I}_2(G)$ is unique and has full support.	
\end{thm}
\begin{proof}By Theorem~\ref{th:mainMLE}, the MLE exists within the convex exponential family $\mathcal{P}_2(G)$. Since $\mathcal{I}_2(G)$ is an exponential subfamily of that, the MLE also exists within $\mathcal{I}_2(G)$.
\end{proof}
In the following we shall develop an algorithm for calculating the MLE in $\mtp$ Ising model.
\subsection{IPS algorithm for computing the MLE}
\label{subsec:IPS} 

The standard IPS algorithm (see \cite{lau96}, page 82) for computing the MLE without the $\mtp$ restriction works by cycling through all pairs $ij\in E$ and optimizing the likelihood function when fixing the values of all canonical parameters associated with variables other than the given pair, namely
\[h^{-ij}:=(h_v, v\in V\setminus\{i,j\}),\quad J^{-ij}:=(J_{uv}, u,v\in V\setminus \{i,j\}).\]
Dually, this corresponds to fitting the mean value parameters associated with $i,j$ to their empirically observed values, i.e.\
\[\mu_i=\bar x_i,\quad \mu_j = \bar x_j, \quad \Xi_{ij}= M_{ij}.\]
If the MLE exists, then this algorithm is known to converge to the MLE (see \cite{lau96}, page 82). 
We next extend this algorithm to $\mtp$ Ising models.

Let $e_{ij}$ denote the empirical distribution of $(X_i,X_j)$. Note that this distribution depends on the sufficient statistics through the formula
$$\begin{aligned}
	e_{ij}(1,1)=(1+\bar x_i +\bar x_j + M_{ij})/4,\quad\qquad
&e_{ij}(1,-1)=(1+\bar x_i -\bar x_j - M_{ij})/4,\\
e_{ij}(-1,1)=(1-\bar x_i +\bar x_j - M_{ij})/4,\quad\qquad 
&e_{ij}(-1,-1)=(1-\bar x_i -\bar x_j + M_{ij})/4.
\end{aligned}$$

We now assume that $e_{ij}(1,-1)>0$ and $e_{ij}(-1,1)>0$ for all $ij\in E$, which ensures that $-1<\bar x_i<1$ for all $i\in V$ and that the MLE has full support; see Theorem~\ref{th:mainMLE}. By Corollary~\ref{cor_MLE_MTP2} and the following paragraph, for edges where $S_{ij}=M_{ij}-\bar x_i\bar x_j< 0$, it holds that $J_{ij}=0$. For the other edges it holds that
\[e_{ij}(1,1)\;\geq\;  (1+\bar x_i +\bar x_j + \bar x_i \bar x_j)/4 \;=\;(1+\bar x_i)(1+\bar x_i)/4\;>\; 0\]
and, similarly, $e_{ij}(-1,-1) \geq (1-\bar x_i)(1-\bar x_i)>0.$

The IPS algorithm is initialized in any point inside the model such as the uniform distribution or the distribution where all variables are mutually independent with mean $\hat \mu=\bar x$. The update for the edge $ij\in E$ can  be expressed as 
\begin{equation}\label{eq:Alg1update}
p(x)\;\leftarrow\; p(x) \frac{e_{ij}(x_i,x_j)}{p_{ij}(x_i,x_j)}= p(x_{-ij}\cd x_i,x_j) e_{ij}(x_i,x_j) =p(x) q_{ij}(x_i,x_j).	
\end{equation}
Using \eqref{eq:logoddsIsing} we easily verify that $J_{ij}$ is the only entry of $J$ affected by this update.   Exploiting that $q_{ij}(x_i,x_j)>0$, we can define
\begin{equation} \label{eq:deltaips}\Delta_{ij}\;:=\;\frac{1}{4} \log \frac{q_{ij}(1,1)q_{ij}(-1,-1)}{q_{ij}(1,-1)q_{ij}(-1,1)}. 
\end{equation}
Using a  mixed parametrization (see~\cite{barndorff:78}) with $(\mu_i,\mu_j, J_{ij})$ and  canonical para- meters for all other indices, the update step can equivalently~be~expressed~as
\[J_{ij} \leftarrow J_{ij}+ \Delta_{ij},\quad \mu_i\leftarrow \bar x_i, \quad \mu_j\leftarrow \bar x_j,\]
where all other entries of $(h,J)$ remain unchanged. 

To ensure the $\mtp$ constraint, it is natural to replace $J_{ij}$ with zero if the update becomes negative and then recalculate $(h_i,h_j)$ to comply with the requirement $(\mu_i,\mu_j)=(\bar x_i,\bar x_j)$.

Alternatively we can express the update in terms of mean value parameters by letting 
$\hat \Xi_{ij}\leftarrow M_{ij}+\lambda^*$. To compute $\lambda^*$ define $e^*_{ij}=e_{ij}(\lambda^*)$  by
\begin{eqnarray*}
e^*_{ij}(1,1)&=&(1+\bar x_i +\bar x_j + \hat \Xi_{ij})/4\;=\;e_{ij}(1,1)+\lambda^*/4,\\
e^*_{ij}(1,-1)&=&(1+\bar x_i -\bar x_j - \hat \Xi_{ij})/4\;=\;e_{ij}(1,-1)-\lambda^*/4,\\
e^*_{ij}(-1,1)&=&(1-\bar x_i +\bar x_j -\hat \Xi_{ij})/4\;=\;e_{ij}(-1,1)-\lambda^*/4,\\
e^*_{ij}(-1,-1)&=&(1-\bar x_i -\bar x_j + \hat \Xi_{ij})/4\;=\;e_{ij}(-1,-1)+\lambda^*/4,
\end{eqnarray*}  and define $q^*_{ij}=e^*_{ij}/p_{ij}$. Then $\lambda^*$ is given by the solution to the equation
\begin{equation}\label{eq_to_solve}
\Delta_{ij}(\lambda)= -J_{ij},
\end{equation}
where
\begin{equation*}  \Delta_{ij}(\lambda^*)\;=\;\frac{1}{4} \log \frac{q^*_{ij}(1,1)q^*_{ij}(-1,-1)}{q^*_{ij}(1,-1)q^*_{ij}(-1,1)}.
\end{equation*}
Note that $\Delta_{ij}(\lambda)$ is strictly increasing in $\lambda$, $\Delta_{ij}(0)< -J_{ij}$, and  $\Delta_{ij}(\lambda) \to \infty$ for $\lambda\to \min(e_{ij}(1,-1),e_{ij}(-1,1))$. Hence there is a unique solution $\lambda^*$ with $\lambda^*>0$. Letting $x=\lambda/4$, then (\ref{eq_to_solve}) becomes
$$
\log\left(\frac{(e_{ij}(1,1)+x)(e_{ij}(-1,-1)+x)}{(e_{ij}(-1,1)-x)(e_{ij}(1,-1)-x)}\right)\;=\;-\log\left(\frac{p_{ij}(1,1)p_{ij}(-1,-1)}{p_{ij}(-1,1)p_{ij}(1,-1)}\right) -4J_{ij},
$$
or equivalently,
$$
\frac{(e_{ij}(1,1)+x)(e_{ij}(-1,-1)+x)}{(e_{ij}(-1,1)-x)(e_{ij}(1,-1)-x)}\;=\;\frac{p_{ij}(1,1)p_{ij}(-1,-1)}{p_{ij}(-1,1)p_{ij}(1,-1)}\cdot e^{-4J_{ij}}.
$$
Denoting the right-hand side of the above equation by $R$, multiplying both sides by $(e_{ij}(-1,1)-x)(e_{ij}(1,-1)-x)$, and moving everything to the left, we obtain a quadratic equation $ax^2+bx+c=0$ with $a=1-R$,
$$
b=e_{ij}(1,1)+e_{ij}(-1,-1)+R(e_{ij}(-1,1)+e_{ij}(1,-1)),
$$
$$ c=e_{ij}(1,1)e_{ij}(-1,-1)-R(e_{ij}(-1,1)e_{ij}(1,-1)).
$$
Hence the solution $\lambda^*=4x^*$ is given by taking $x^*$ to be the positive root of this quadratic equation.  
Using $\lambda^*$ we can then update $p(x)$ as follows:
\begin{equation}\label{eq:Alg1update_star}
p(x)\;\leftarrow\; p(x) \frac{e^*_{ij}(x_i,x_j)}{p_{ij}(x_i,x_j)}.
\end{equation}

\noindent The full procedure is presented in Algorithm~\ref{algthm:IPS}. In the following theorem we show that this procedure indeed converges to the MLE (if it exists).

\begin{algorithm}[!t]
\caption{IPS-type algorithm for computing the MLE in $\mtp$ Ising models.}
\label{algthm:IPS}
\begin{algorithmic}
\vspace{0.2cm}
\State {\bf input:\;\;\;}
Sample moments  $(\bar x, M)$, a graph $G=(V,E)$,  and precision $\epsilon$.

\State {\bf output:} The MLE $(\hat p, \hat G, \hat \mu, \hat\Xi)$. 
\State \rule{5.75in}{0.1mm}

\vspace{0.2cm}

\State {\bf initialize} $\mu= \bar x$; $p(x) = 2^{-|V|}\prod_{v\in V}(1-(-1)^{x_v}\mu_v) \textrm{ for all } x\in \cX$; $\Xi=\mathbf{I}$;
\State {\bf initialize} $E^+=\{uv\in E \mid M_{uv}> \bar x_{u}\bar x_{v}\}$; $\hat E=\emptyset$;

\State

\Repeat
\For{$ij\in E^+$}
\State \textbf{calculate} $\Delta_{ij}$ by (\ref{eq:deltaips});
\State \textbf{calculate} $J_{ij}$  by (\ref{eq:logoddsIsing});
\If{$\Delta_{ij} + J_{ij} > 0$}
\State \textbf{update} $p$ by (\ref{eq:Alg1update});
\State $\hat E \leftarrow \hat E \cup \{ij\}$;
\Else 
\State \textbf{solve} $\Delta_{ij}(\lambda)=-J_{ij}$; 
\State \textbf{update} $p$ by (\ref{eq:Alg1update_star}); 
\State $\hat E \leftarrow\hat E \setminus \{ij\}$;
\EndIf
\EndFor
\State \textbf{calculate} $(\mu,\Xi)$ from $p$;
\Until {$\max_{v\in V}|\hat \mu_{v}- \bar x_{v}|<\epsilon$ \textbf{ and } $\Xi\geq M$ \textbf{ and } $\max_{uv\in \hat E}|\Xi_{uv}-M_{uv}|<\epsilon$;}
\State
\Return $p,\; \hat G=(V,\hat E),\; \mu,\; \Xi$. 

\end{algorithmic}
\end{algorithm}

\begin{thm}If the MLE for the $\mtp$ Ising model on the undirected graph $G=(V,E)$ exists, then the output of Algorithm~\ref{algthm:IPS} converges to the MLE for $\epsilon\to 0$.
\end{thm}
\begin{proof} Let $(h,J)$ denote the canonical parameters of the exponential family. 
	Then the log-likelihood function satisfies 
\[	-\frac{1}{n}\log L(h,J)= \log c(h,J) - h^T \bar x - \tr(JM)/2,\]
where $c(h,J)$ is the normalizing constant of the exponential family.
 We fix a value $(h^0,J^0)$ with $J^0_{uv}\geq 0$ and consider the following restricted convex optimization problem:
\begin{equation*}
\begin{aligned}
& \underset{(h,J)}{\text{minimize}}
& & \log c(h,J) - h^T \bar x - \tr(JM)/2\\
& \text{subject to} && J_{ij}\geq 0, \; h_u=h^0_u, u\in V\setminus\{i,j\},\; J_{uv}= J^0_{uv} \text{ for $uv\neq ij$.}
\end{aligned}
\end{equation*}
\mbox{Exploiting that most entries of $(h,J)$ are fixed, this problem is equivalent to}
\begin{equation*}
\begin{aligned}
& \underset{(h_i,h_j,J_{ij})}{\text{minimize}}
& & \log c(h,J) - h_i\bar x_i - h_j\bar x_j - J_{ij}M_{ij}\\
& \text{subject to} && J_{ij}\geq 0,
\end{aligned}
\end{equation*}
where the fixed values $h_u=h^0_u$, $u\in V\setminus\{i,j\}$ and  $J_{uv}= J^0_{uv}$  for $uv\neq ij$ enter into the function $\log c(h,J)$. Since also this subfamily is a convex exponential family, the solution to this optimization problem is uniquely determined by:
\begin{itemize}
	\item[(i)] Primal feasibility:\; $\hat J _{ij}\geq 0$
	\item[(ii)] Dual feasibility:\; $\hat \mu_i=\bar x_i $, $\hat \mu_j = \bar x_j$,  and $\hat \Xi_{ij}\geq M_{ij}$,
	\item[(iii)]  Complementary slackness:\; $(\hat \Xi_{ij}-M_{ij})\hat J _{ij}=0$. 
\end{itemize}
Thus, if $J^0_{ij}+\Delta_{ij}\geq 0$ we update as in (\ref{eq:Alg1update}). Else we update as in (\ref{eq:Alg1update_star}). 

Note that every step of the algorithm maximizes the likelihood over a section. In addition, any fixed point of the algorithm satisfies the conditions in Corollary~\ref{cor_MLE_MTP2} and hence must be equal to the unique MLE. Furthermore, the updates depend continuously on $p$. Hence the algorithm is an instance of iterative partial maximization as described in \cite[page 230]{lau96} and is therefore convergent with the unique MLE as limit. 
\end{proof}

We note a computational issue with Algorithm~\ref{algthm:IPS}. As stated above, the algorithm requires visiting all possible states $x\in \cX$, which becomes computationally prohibitive for large $d$ as the computational effort is then exponential in $d$. This problem can be overcome by an appropriate use of probability propagation as in \cite{jirousek:preucil:95}.  More precisely, instead of representing $p$ by its \emph{values} $p(x), x\in \cX$, we represent $p$ by a set of \emph{potentials} $\psi_{ij}, ij\in E$, such that
\[p(x) \propto \prod_{ij\in E}\psi_{ij}(x_i,x_j) = \prod_{ij\in E} \exp(x_ix_jJ_{ij}).\]
Whenever a marginal $p(x_u,x_v)$ is required for the update, it is calculated from $J$ by probability propagation as e.g.\ described in \cite{cowell:etal:99}. Then instead of updating $p$ itself, the update (\ref{eq:Alg1update}) or (\ref{eq:Alg1update_star}) is performed by updating $J$ only.  This reduces the computational effort  to become linear in the maximal clique size of $G$ rather than $d$.

Finally, note that since the algorithm runs entirely in terms of probabilities $p(x)$, a simple modification  of the algorithm as in \cite{lauritzen:02,lau96}  guarantees convergence even when the MLE does not exist within the exponential family. We refrain from providing the details of this modification.

\begin{ex}\label{ex:mouss.alg}
	Consider again the data in Example~\ref{ex:moussouris}. On this data, Algorithm~\ref{algthm:IPS} converges in one step and the maximum likelihood distribution is given by a rational function of the data. The corresponding MLEs  are
$$
\hat\Sigma\;=\;\begin{bmatrix}
1 & 0.5 & 0.25 & 0.125\\
0.5 & 1 &0.5 &0.25\\
0.25 &0.5 & 1 & 0.5\\
0.125 &0.25 & 0.5 & 1\end{bmatrix},\qquad \hat J\;=\;\frac{\log(3)}{2}\begin{bmatrix}
0 & 1 & 0 & 0\\
1 & 0 &1 &0\\
0 &1 & 0 & 1\\
0 &0 & 1 & 0\end{bmatrix}
$$
This is a very special example, where the following three MLEs all coincide:
\begin{enumerate}
	\item MLE computed under $\mtp$ for general binary distributions. 
	\item MLE computed for the $\mtp$ Ising model over the complete graph.
	\item MLE computed for the Ising model over the chain graph in Figure~\ref{fig:4chain}.
\end{enumerate}
The equivalence of (2) and (3) follows from Corollary~\ref{cor_polyhedral} whereas (1) and (2) are usually not equivalent. \hfill\qed 
\end{ex}

\subsection{Totally positive Ising models with no external field}
\label{subsec:noext}

A special example of a symmetric binary distribution is the Ising model with no external field, that is, a family of binary distributions over $\cX=\{-1,1\}^d$ of the form
\begin{equation}\label{eq:ising}
p(x)\;\;=\;\;\frac{1}{c(J)}\exp\left(x^T J x/2\right).
\end{equation}
This was termed \emph{the palindromic Ising model} in \cite{marchetti:wermuth:16}. 
The space of canonical parameters is the set $\S^d_0$ of all symmetric $d\times d$ matrices with $0$ in the diagonal. The mean parameter is  $\Sigma=\Xi=\E X X^T$, which is the correlation matrix because $\Sigma_{ii}=\E X_i^2=1$ and $\E X_i=0$. By Proposition \ref{prop:mtpquadratic}, the quadratic exponential family is $\mtp$ if and only if $J_{ij}\geq 0$ for all $i\neq j$. In \cite{marchetti:wermuth:16} these models have been studied  as a close proxy to the Gaussian distribution since (\ref{eq:ising}) becomes almost identical to the Gaussian density by letting $J=-K$ in this expression.

As a consequence of  Proposition~\ref{prop:symcounts}, we note that Algorithm~\ref{algthm:IPS} also converges for palindromic Ising models by working with the symmetrized sample $U+U^s$. However,  the algorithm can be simplified using
\begin{eqnarray*}
e_{ij}(1,1)&=&e_{ij}(-1,-1)=(1+ M_{ij})/4\\
e_{ij}(1,-1)&=&e_{ij}(-1,1)=(1 - M_{ij})/4.
\end{eqnarray*}
In addition, 
\begin{equation} \label{eq:deltaips_sym}\tilde \Delta_{ij}(\lambda)=\frac{1}{2}\log \frac{p_{ij}(-1,1)(1+M_{ij}+\lambda)}{p_{ij}(1,1)(1-M_{ij}-\lambda)}.\end{equation} 
can be used to determine $\lambda$ to ensure the $\mtp$ property is preserved under the update. We refrain from giving the full details of the simplified steps in this algorithm.

\section{Application to psychological disorders}
\label{sec:application}

In this section, we illustrate the developed methods via a real data case study. We analyze data  obtained from the National Comorbidity Survey Replication study \cite{alegria,kessler2004us} (NCS-R data), which was also analyzed in \cite{borsboom2013network}. The data consists of 9282 observations of 18 binary variables, namely \texttt{depr} (Depressed mood), \texttt{inte} (Loss of interest), \texttt{weig} (Weight problems), \texttt{mSle} (Sleep problems), \texttt{moto} (Psychomotor disturbances), \texttt{mFat} (Fatigue), \texttt{repr} (Self reproach), \texttt{mCon} (Concentration problems), \texttt{suic} (Suicidal ideation), \texttt{anxi} (Chronic anxiety/worry), \texttt{even} (Anxiety $>1$ event), \texttt{ctrl} (No control over anxiety), \texttt{edge} (Feeling on edge), \texttt{gFat} (Fatigue), \texttt{irri} (Irritable), \texttt{gCon} (Concentration problems), \texttt{musc} (Muscle tension), \texttt{gSle} (Sleep problems). These variables are symptoms related to two disorders, namely major depression (MD) and generalized anxiety disorder (GAD). The symptoms that are known to appear in both disorders are sleep problems, fatigue, and concentration problems. These so-called bridge variables appear in pairs  \texttt{mSle}, \texttt{gSle}, \texttt{mFat}, \texttt{gFat} and \texttt{mCon}, \texttt{gCon}. 

The contingency table resulting from this dataset is very sparse with only 872 out of 65536 elementary events observed; 5667 out of 9282 respondents recorded none of the listed symptoms. All variables are positively correlated in the sample. Although the sample distribution is not $\mtp$, assuming total positivity is justified in this  application, since the symptoms are likely to appear jointly.  The sample does not satisfy  the conditions of Theorem~\ref{thm:fullMLE}, because the variables \texttt{anxi}  and \texttt{even} are perfectly correlated with each other and with seven other variables in the sample distribution. For the analysis, we therefore removed these two problematic variables and ran Algorithm~\ref{algthm:IPS} on the remaining contingency table of size $2^{16}$. We used a convergence criterion of $\epsilon=10^{-4}$. The algorithm converged after 28 iterations through all 120 variable pairs which in our current rough  implementation took 37 minutes on a laptop.   We note that using the algorithm and software developed in \cite{forcina2000} fitting the unconstrained $\mtp$ model failed due to space limitations. 

\begin{figure}[!t]
	\includegraphics[scale=.6]{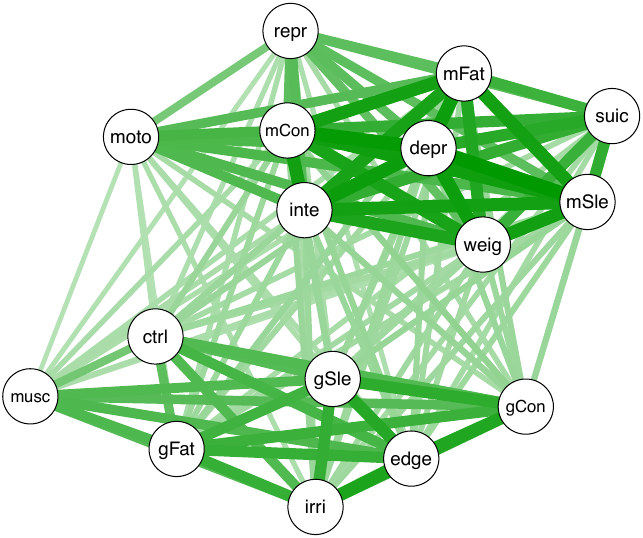}\includegraphics[scale=.6]{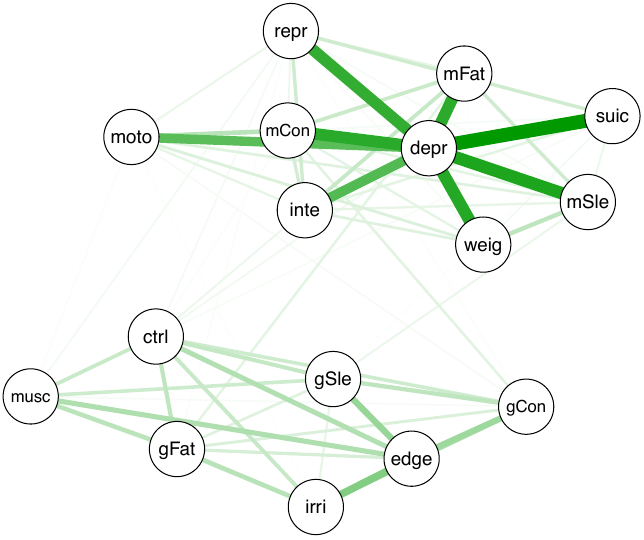}
	\vspace{-0.6cm}
	\caption{(left) Sample correlation network for the NCS-R data, (right) the corresponding network of $\hat J$.}\label{fig:Jdep}
\end{figure}

Figure~\ref{fig:Jdep} shows the network corresponding to the sample correlation matrix (left) and the MLE $\hat J$ (right). The magnitude of an entry $ij$ in the matrix is represented by the thickness of the corresponding edge. The sample correlation network including the two nodes \texttt{anxi} and \texttt{even} is also shown in Figure~2b of \cite{borsboom2013network}. The sparsity of the MLE $\hat{J}$ as compared to the sample correlation matrix is striking; it contains 72 edges as compared to 120 edges in the complete graph on 16 vertices. In addition, the graphical model given by $\hat{J}$ cleanly separates into two blocks with the upper block prominently containing a star graph with center \texttt{depr}. This resembles Figure~4 in \cite{borsboom2013network}, where this subgraph is called a causal skeleton of the covariance graph and was obtained based on rankings by 12 clinicians.  
Moreover, we note that the bottom block, while less prominently, also contains a star graph centered at the variable \texttt{edge}. Finally, note that the three most significant edges across the two blocks are between pairs of bridge variables. This analysis shows that Algorithm~\ref{algthm:IPS} resulted in an interpretable sparse graphical model with a network that seems relevant for the application.  

The graphical model learned by Algorithm~\ref{algthm:IPS} fits reasonably well: The value of the log-likelihood function at the MLE is {-28{,}767.3},  while the value of the log-likelihood function of the unrestricted Ising model (fitted using the \texttt{loglin} function in \textsc{R}) is {-28{,}682.45}. This results in a likelihood ratio statistic of {169.7} which appears high compared to a $\chi^2$ distribution with $120-72=48$ degrees of freedom. However, the exact and asymptotic distributions of this statistic are unknown; the asymptotic distribution is a mixture of $\chi^2$-distributions with different degrees of freedom, but with unknown weights. 

We also calculated the split-likelihood ratio test statistic as described in \cite{wasserman:etal:19} and this resulted in a test statistic of
$U_n=1.8\times10^{-58}$ which does not reject the $\mtp$ hypothesis for any level $\alpha$ as it should be compared to $1/\alpha$. Hence it appears that the $\mtp$ analysis of this dataset is appropriate.

\section*{Acknowledgements}

We would like to thank Antonio Forcina for making his {\sc Matlab} code from \cite{forcina2000} available to us. We have also benefited from discussions with B\'{e}atrice de Tili\`{e}re. This research was supported through the program ``Research in Pairs'' by the Mathematisches Forschungsinstitut Oberwolfach in 2018.  Caroline Uhler was partially supported by NSF (DMS-1651995), ONR (N00014-17-1-2147 and N00014-18-1-2765), IBM, and a Simons Investigator Award. Piotr Zwiernik was supported by the Spanish Ministry of Economy and Competitiveness (MTM2015-67304-P), Beatriu de Pin\'{o}s Fellowship (2016 BP 00002), and the program Ayudas Fundaci\'{o}n BBVA (2017).


\end{document}